\documentclass{pasj00}
\onecolumn

\begin{document}
\SetRunningHead{Takahashi et al.}{Evolution of Galaxy Clusters}
\Received{2001 June 22}
\Accepted{2001 October 11}

\title{Collisional Evolution of Galaxy Clusters \\
and the Growth of Common Halos}

\author{
  Koji \textsc{Takahashi}\altaffilmark{1},
  Tomohiro \textsc{Sensui}\altaffilmark{1},
  Yoko \textsc{Funato}\altaffilmark{2},
  \and
  Junichiro \textsc{Makino}\altaffilmark{1}}
\altaffiltext{1}{Department of Astronomy, School of Science,
                 The University of Tokyo,\\ Bunkyo-ku, Tokyo 113-0033}
\email{takahasi@astron.s.u-tokyo.ac.jp}
\altaffiltext{2}{General Systems Sciences, Graduate Division of
                 International and Interdisciplinary Studies,\\
                 The  University of Tokyo, Meguro-ku, Tokyo 153-8902}


\maketitle

\begin{abstract}
We investigated the dynamical evolution of clusters of galaxies in 
virial equilibrium using Fokker--Planck models and self-consistent
$N$-body models. In particular, we focused on the growth of a common halo,
which is a cluster-wide halo formed by matter stripped from galaxies,
and the development of a central density cusp. The Fokker--Planck
models include the effects of two-body gravitational encounters both
between galaxies and between galaxies and common halo particles.
The effects of tidal mass stripping from the galaxies due to close
galaxy--galaxy encounters and accompanying dissipation of the orbital
kinetic energies of the galaxies were also taken into account in the
Fokker--Planck models. We find that the results of
the Fokker--Planck models are in
excellent agreement with those of the $N$-body models regarding the
growth of the common halo mass and the evolution of the cluster density
profiles. In the central region of the cluster, a shallow density cusp,
approximated by $\rho (r) \propto r^{-\alpha}$ ($\alpha \sim$ 1), develops.
This shallow cusp results from the combined effects of two-body
relaxation and tidal stripping. The cusp steepness, $\alpha$, weakly
depends on the relative importance of the tidal stripping. When the effect
of stripping is important, the central velocity dispersion decreases
as the central density increases and, consequently, a shallow
($\alpha <2$) cusp is formed. In the limit of no stripping, usual
gravothermal core collapse occurs, i.e. the central velocity
dispersion increases as the central density increases with
a steep ($\alpha >2$) cusp left.
We conclude
from our consideration of the origin of the cusp demonstrated here that
shallow cusps should develop in real galaxy clusters.
\end{abstract}

\section{Introduction}\label{sec:intro}

Many observations suggest that clusters of galaxies
contain substantial amounts of mass in the form of
a background intergalactic stellar component 
(e.g., V\'\i lchez-G\'omez et al.\ 1994;
Scheick, Kuhn 1994; Ferguson et al.\ 1998 and references therein).
It has usually been thought that
such a component has its origin in stars stripped from galaxies
by galaxy--galaxy encounters, or by the cluster tidal field.
However, detailed numerical simulations of galaxy clusters
are required to confirm how efficient such mass stripping is
and how the structure of the background component evolves.

It was only recently that fully self-consistent $N$-body simulation
of galaxy clusters,
with each galaxy represented by many particles, was made possible,
because such a full simulation requires huge computing resources.
Funato et al.\ (1993) investigated the evolution of isolated 
clusters of galaxies using self-consistent $N$-body simulations.
Similar simulations were also performed by Bode et al.\ (1994) and
Garijo et al.\ (1997).
More recently, Sensui et al.\ (1999, SFM99; 2000, SFM00)
followed up the study of Funato et al.\ (1993) by performing
larger $N$ simulations for a wider variety of initial conditions.
In their simulations, clusters were initially in virial equilibrium,
and all of the cluster mass was initially bound to individual galaxies;
a cluster was composed of about 100 identical galaxies, and
each galaxy was modeled with 500--4000 particles.
SFM99 and SFM00 found that, in all cases,
more than half of of the total mass is stripped from galaxies
in a few crossing times of the cluster,
and that the stripped matter forms a smooth cluster-wide common halo.
They also found that a density cusp, approximated as $\rho \propto
r^{-\alpha}$ ($\alpha$ being form $-1$ to $-1.5$),
develops in the central region.
However, the physical mechanism of the cusp formation
was not clear.

Even if we are allowed to 
regard clusters of galaxies as ``pure'' self-gravitating
many-body systems composed of point masses (stars and dark matter particles),
we have to admit that their evolution is complicated by many physical
processes which take place simultaneously:
gravitational encounters, physical collisions, and merging between galaxies;
galaxy--background particle interactions,
mass stripping from galaxies, internal evolution of galaxies, etc.
Therefore, there is no doubt that fully self-consistent $N$-body simulations
are most desirable to obtain qualitatively and quantitatively
reliable results on cluster evolution
(see Garc\'\i a-G\'omez et al.\ 1996).

Nevertheless, more idealized or more simplified models that include (part of)
relevant physical processes in approximate ways,
such as Fokker--Planck (FP, hereafter) models which we used in the
present study, are still very useful regarding two points.
First, we can quickly obtain
an overview of cluster evolution based on various initial conditions
by using approximate models, which are inexpensive in computation time.
Second, it is expected that
we can understand more clearly the role and importance
of each physical process,
because we can easily separate the effects of various processes
in approximate models.
On the other hand, it is often difficult to ``understand''
the results of $N$-body simulations,
where all processes are automatically included
and connected through a complicated procedure.
Thus, $N$-body and approximate models are complementary to each other,
and comparison between these models
will be particularly helpful to our further understanding
of the physics of cluster evolution,
regardless of whether agreement or disagreement is found.

Merritt (1983, 1984, 1985) explored the evolution of clusters of galaxies
using such approximate models based on an isotropic orbit-averaged
FP equation (Cohn 1980).
The FP equation describes the evolution of
the phase-space distribution function driven by gravitational two-body encounters,
and has been applied mainly to globular clusters
(see, e.g., Spitzer 1987 for a review).
A typical cluster of galaxies is composed of $N=$ 100--1000 galaxies,
and thus the time scale of two-body relaxation between galaxies, $T_{\rm r}$,
is of the same order as
the crossing time of the cluster, $T_{\rm cr}$
($T_{\rm r}/T_{\rm cr} \approx 0.1 N/\ln N$).
Therefore, two-body relaxation is expected to play an important role
in the evolution of galaxy clusters.
In his models, Merritt also took account of mass stripping from galaxies
due to tidal interactions between galaxies.
Since the stripped particles (stars or dark matter particles) were 
considered to be much lighter than galaxies, these particles
were treated as being massless.
Dynamical friction between the galaxies and the background 
massless particles was also included.
Thus, in Merritt's models the evolution of the galaxies and
the background was treated self-consistently.
FP models are continuum models, and therefore have an advantage
that they are free from statistical fluctuations,
which are unavoidable in $N$-body models and make it difficult
to understand the underlying physics.
Yepes et al.\ (1991) and Yepes and Dom\'\i nguez-Tenreiro (1992)
studied the evolution of galaxy clusters by solving the moment equations
of the FP equation. They also included a background component of
massless particles in their models,
but did not take account of tidal stripping.

The purpose of the present work was to investigate
the dynamical evolution of clusters of galaxies after virialization.
In particular, we aimed to clarify the physical mechanisms 
of the growth of the common halo,
and of the development of the central density cusp, 
which were investigated by SFM99 and SFM00 with $N$-body models.
For this purpose, we used FP models, which are
based on Merritt's (1983) formulation, but are improved in many aspects,
and compare them with the self-consistent $N$-body models
of SFM99 and SFM00.
As shown below, the results of the FP and $N$-body models
agree with each other surprisingly well.
This suggests that our FP models correctly include all important
physical processes, and thus can be used as a useful tool for studying
the dynamics of galaxy clusters.

This paper is organized as follows.
The basic equations of the FP models are given in section 2.
Section 3 describes the initial conditions of our simulations,
and section 4 briefly describes the numerical schemes used for our
FP and $N$-body simulations.
The simulation results are given in sections 5 and 6;
in section 5, the FP and $N$-body models are compared
with respect to the growth of the common halo and the development
of the central cusp;
in section 6, the physical mechanism of the cusp formation
is examined in detail based on FP models.
We summarize and discuss our results in section 7.

\section{Fokker--Planck Models}\label{sec:fpmodel}

In this section, we describe FP models of clusters of galaxies.
In FP models, individual galaxies are treated as point particles,
except that the effects of tidal stripping are considered.
Detailed FP simulations of galaxy clusters were
first performed by Merritt (1983).
We closely follow his formulation.
There are, however, three major differences between his and our models:
(1) Merritt used isotropic FP models, but we used anisotropic FP models.
(2) We used a new cross-section formula of mass stripping from galaxies
based on recent $N$-body experiments by Funato and Makino (1999, FM).
(3) Dissipation of the orbital kinetic energies of galaxies during
galaxy--galaxy encounters was also considered, while
it was neglected in Merritt's simulations.

\subsection{The Fokker--Planck Equation}\label{sec:fpmodel1}

Consider a self-gravitating many-body system
composed of different mass components with particle mass $m_i$
($i=0,1,2,\dots$).
We write the {\it number density} in $\mu$-space, i.e. the distribution
function, of component $m_i$ as $f_i(\mbox{\boldmath $r$},\mbox{\boldmath $v$},t)$.
By assuming that
the system is spherically symmetric and in dynamical equilibrium,
$f_i$ can be written as a function of the energy per unit mass, $E$,
and the angular momentum per unit mass, $J$.
Then, the orbit-averaged FP equation, under the fixed gravitational potential,
can be written as follows (Cohn 1979; Takahashi 1997):
\begin{eqnarray}
A \frac{\partial f_i}{\partial t}
&=& \frac{\partial}{\partial E}\left( D_{E i} f_i
+ D_{EE i} \frac{\partial f_i}{\partial E} 
+ D_{ER i}\frac{\partial f_i}{\partial R} \right) \nonumber \\
&+&  \frac{\partial}{\partial R} \left( D_{R i} f_i 
+ D_{RE i}\frac{\partial f_i}{\partial E} 
+ D_{RR i}\frac{\partial f_i}{\partial R} \right) \,, \label{eq:FP}
\end{eqnarray}
where $R$ denotes the scaled angular momentum,
$A$ is a weight function, and
$D_{E i}, D_{EE i}$, etc. are flux coefficients.
See Takahashi (1997)
for a complete description of this equation.
To follow the evolution of the system, 
this FP equation is solved in combination with
Poisson's equation (Cohn 1979).

We introduce the {\it mass density} in phase space,
\begin{equation}
g_i \equiv m_i f_i .
\end{equation}
Here, we choose the $i=0$ component for representing the background,
or a common halo.
We can assume that the mass of a background particle (star) is negligible
compared to that of a galaxy.
However, the mass density of background particles is comparable to 
that of galaxies.
We now take the limit $m_0 \to 0$ while keeping $g_0$ finite
(i.e. $f_0 \to \infty$).
In the FP equation (\ref{eq:FP}), component mass $m_i$ appears
only in the flux coefficients.
In the limit  $m_0 \to 0$,
the expressions for the first and second-order flux coefficients look like
\begin{eqnarray}
D_{E 0} &=&0 ,\quad
D_{E i} \propto m_i \sum_{{\rm all}\; j} g_j \quad (i=1,2,3,\dots),
\\
D_{EE i},\ D_{ER i},\ D_{RE i},\ D_{RR i} &\propto& \sum_{j \ne 0} m_j g_j  \quad (i=0,1,2,\dots).
\end{eqnarray}
Note that we write here only the dependence on $m_i$ and $g_i$ symbolically.
Exact expressions can be easily obtained from the equations given
in the appendix of Takahashi (1997).

The FP equation for the background component ($i=0$) becomes
\begin{equation}
A \frac{\partial g_0}{\partial t} =
\frac{\partial}{\partial E} \left( D_{EE 0} \frac{\partial g_0}{\partial E}
+ D_{ER 0}\frac{\partial g_0}{\partial R} \right)
+\frac{\partial}{\partial R} \left( D_{RE 0}\frac{\partial g_0}{\partial E}
+D_{RR 0}\frac{\partial g_0}{\partial R} \right)
\label{eq:FP0} ,
\end{equation}
which is obtained
by multiplying equation (\ref{eq:FP}) by $m_0$ and then taking the
limit $m_0 \to 0$.
Since the background particles are massless,
they are completely collisionless, i.e.,
they do not interact with each other.
However,
galaxies suffer dynamical friction from the background
particles that have a finite mass density. Thus, the orbital energy
of the galaxies is transferred to the background.

\subsection{Mass Stripping Rates}\label{sec:fpmodel2}

Consider an encounter between 
{\it a test galaxy} of mass $m$ and {\it a field galaxy} of mass $m_f$,
with impact parameter $p$ and relative velocity $V$.
FM showed, both numerically and analytically, that the
relative change of the test galaxy's mass
in an encounter can be approximated by
\begin{equation}
\frac{\Delta m}{m} = - C' \left(\frac{r_{\rm g}}{p}\right)^{\eta_r}
                       \left(\frac{v_{\rm g}}{V}\right)^{\eta_v}
                       \left(\frac{m_f}{m}\right)^{\eta_m},
\label{eq:ml0}
\end{equation}
with $\eta_r \simeq$ 2--3 and $\eta_v \simeq$ 2--3,
for the cases with $p \gtsim r_{\rm g}$, $V \gtsim v_{\rm g}$, and $m= m_f$.
Here, $C'$ is a dimensionless constant,
and $r_{\rm g}$ and $v_{\rm g}$ are the size (represented by the
virial radius in the following) and
the three-dimensional internal velocity dispersion of the test galaxy,
respectively.
Since FM did not carry out simulations of encounters between
unequal-mass galaxies, the value of $\eta_m$ is not determined from
their simulations.
In the impulse approximation,
the velocity change of a star, $\Delta v$, is proportional to $m_f$. 
Therefore, if the effect of the term $(\Delta v)^2$ is dominant
in removing mass from the test galaxy, we may expect $\eta_m \simeq 2$;
on the other hand, if the effect of the term $v \Delta v$ is dominant,
$\eta_m \simeq 1$ (see Merritt 1983; FM).

Now suppose that a test galaxy experiences successive encounters with
field galaxies of mass $m_f$.
The rate of change in the test galaxy's mass is given by
\begin{equation}
\frac{1}{m}\frac{dm}{dt} = n_f \int_{p_{\rm min}}^{p_{\rm max}} 2\pi p dp
\int F(\mbox{\boldmath $V$})d^3\mbox{\boldmath $V$} V \frac{\Delta
m}{m} ,
\label{eq:ml1}
\end{equation}
where $n_f$ is the number density of the field galaxies
and $F(\mbox{\boldmath $V$})$ is the (normalized) distribution
function of the relative velocity between the test and field galaxies;
$p_{\rm min}$ and $p_{\rm max}$ are the minimum and maximum impact parameters,
respectively.
If we assume a Maxwellian distribution with a three-dimensional velocity
dispersion, $V_f$, for $F$, this equation becomes
\begin{equation}
\frac{1}{m}\frac{dm}{dt} =
-\left(\frac{8\pi}{3}\right)^{1/2} C' n_f r_{\rm g}^2 V_f
\left(\frac{v_{\rm g}}{V_f}\right)^{\eta_v}
\left(\frac{m_f}{m}\right)^{\eta_m}
\int_{x_{\rm min}}^{x_{\rm max}} x^{1-\eta_r} dx
\int_0^\infty y^{3-\eta_v} \exp(-y^2/2) dy.
\label{eq:ml2}
\end{equation}
Here, we define $x=p/r_{\rm g}$ and $y=\sqrt 3 V/V_f$.
The first integral becomes $\ln (x_{\rm max}/x_{\rm min})$
for $\eta_r=2$, and $1/x_{\rm min}-1/x_{\rm max}$ for $\eta_r=3$.
Therefore, if we take $p_{\rm min}$ and $p_{\rm max}$ as the size
of the galaxy and that of the cluster,
we may consider that the integral remains roughly constant during
the cluster evolution for $\eta_r =$ 2--3.
The second integral is also a constant of the order of unity
for $\eta_{\rm v}=$ 2--3; it equals 1 when $\eta_v=2$, and $(\pi/2)^{1/2}$ 
when $\eta_v=3$.

Therefore, we may rewrite equation (\ref{eq:ml2}) as
\begin{equation}
\frac{dm}{dt} =
-C \rho_f r_{\rm g}^2 V_f
\left(\frac{v_{\rm g}}{V_f}\right)^{\eta_v}
\left(\frac{m_f}{m}\right)^{\eta_m-1} ,
\label{eq:ml3}
\end{equation}
where $C$ is a dimensionless constant factor and
$\rho_f=m_f n_f$ is the mass density of the field
galaxies.
Note that the definition of $C$ depends on $\eta_r$ and $\eta_v$.
When $\eta_v=2$, for example,
$C=(8\pi/3)^{1/2} \ln(p_{\rm max}/p_{\rm min})C'$ for $\eta_r=2$,
and $C=(8\pi/3)^{1/2} (1/p_{\rm min} -1/p_{\rm max})C'$ for $\eta_r=3$.

Equation (\ref{eq:ml3}) includes two parameters of galaxy's internal
structure, $r_{\rm g}$ and $v_{\rm g}$. Because FP models do not
follow the internal evolution of galaxies, we have to adopt some
relations to express the changes in these parameters by $m$ only.
One such relation should be the virial theorem,
$v_{\rm g}^2 \propto Gm/r_{\rm g}$.
As the other relation we assume
$v_{\rm g} \propto m^{\zeta}$ with $\zeta \simeq$ $1/4$--$1/3$ 
(see FM and SFM99).
Using these relations we may write the two parameters as
\begin{equation}
\frac{v_{\rm g}}{v_{\rm g0}}=\left(\frac{m}{m_{\rm g0}}\right)^{\zeta}, \quad
\frac{r_{\rm g}}{r_{\rm g0}}=\left(\frac{m}{m_{\rm g0}}\right)^{1-2\zeta}.
\label{eq:vg-m}
\end{equation}
Finally, we obtain
\begin{equation}
\frac{dm}{dt}(r) =
-C \left(\frac{r_{\rm g0}^2 v_{\rm g0}^{\eta_v}}{m_{\rm g0}^{2-(4-\eta_v)\zeta}}\right)
m^{3-(4-\eta_v)\zeta-\eta_m}
\sum _f m_f^{\eta_m-1}\rho_f(r)V_f^{-(\eta_v-1)}(r) ,
\label{eq:ml4}
\end{equation}
where a sum over all galaxy components $f$ is taken.
We then orbit-average (see, e.g., Takahashi 1997)
equation (\ref{eq:ml4}) to obtain
the mass-loss rate as a function of $E$ and $J$.

\subsection{Energy Dissipation Rates}\label{sec:fpmodel3}

Stars escape from galaxies via galaxy--galaxy encounters.
This is possible
because part of the orbital kinetic energies of galaxies is converted
into the energy of internal motions of the stars in the galaxies.
That is, these encounters are inelastic.
If the sum of the galaxy binding energies is much smaller than
the cluster binding energy,
i.e., if $(v_{\rm g}/V_{\rm cl})^2 \ll 1$ ($V_{\rm cl}$ is the cluster
velocity dispersion),
the effect of energy dissipation due to
inelastic encounters is not very important.
However, for our model clusters shown below, this condition 
is not well satisfied [$(v_{\rm g}/V_{\rm cl})^2 =5/32$],
and therefore energy dissipation is expected to play a non-negligible role.

The binding energy of a galaxy (in virial equilibrium) is
$ \varepsilon_{\rm g} = mv_{\rm g}^2/2$. The rate of its change is
therefore given by
\begin{equation}
\frac{d\ln \varepsilon_{\rm g}}{dt} = \frac{d\ln m}{dt}+\frac{d\ln v_{\rm g}^2}{dt}
= (1+2\zeta) \frac{d\ln m}{dt} < 0 \label{eq:ec} .
\end{equation}
Here, in the second equality, we used equation (\ref{eq:vg-m}).
The mass-loss rate is given by equation (\ref{eq:ml4}).

The decrease in the binding energy of a galaxy should be compensated by
a decrease in the orbital kinetic energies of 
the galaxies involved in encounters with that galaxy.
For simplicity,
we assume that the change in the binding energy of a galaxy
is just equal to the change in its orbital energy,
and that the stripped stars have the same orbital energy per unit mass
as that of the parent galaxy.
This is valid at least on average.
The dissipation rate of the orbital energy of galaxy $m$ is given by
\begin{equation}
m\left(\frac{dE}{dt}\right)_{\rm dis}
= \frac{d\varepsilon_{\rm g}}{dt}
= (1+2\zeta) \frac{\varepsilon_{\rm g}}{m} \frac{d m}{dt}
\label{eq:edis}
\end{equation}
(remember $E$ represents the energy per units mass).
The above assumption ensures that
the total energy of a galaxy cluster, including the internal energies of
the galaxies, is conserved.

\subsection{Implementation of Mass Stripping and Energy Dissipation}
\label{sec:fpmodel4}

The distribution function evolves
owing to mass stripping from galaxies as well as owing to two-body relaxation.
The effect of mass stripping can be included in the FP equation
by adding a source (or sink) term
(see, e.g., Quinlan, Shapiro 1989).
Thus, in this case the FP equation becomes
\begin{equation}
A \frac{\partial f_i}{\partial t} = \Gamma_i + S_i ,
\end{equation}
where $\Gamma_i$ denotes the FP collision term [the right-hand side of equation
(\ref{eq:FP})] and $S_i$ represents the source term due to
mass stripping.
We neglect merging events between galaxies for simplicity.
For the initial conditions which we adopted in this work,
$N$-body simulations showed that mergers were rare (SFM99; SFM00).
Therefore, neglect of merging events would not cause serious errors
for our models.
The source term is calculated in practice as follows.

For each galaxy component $i$ and for each grid point in $(E,R)$ space,
we can
calculate the change in mass $\Delta m_i (E,R)$ during a time step $\Delta t$
from the orbit-averaged stripping rate, $\langle dm/dt \rangle_{\rm OA}$.
Thus, we know that
galaxies of mass $m=m_i$ at time $t$ and position $(E,R)$ will have mass
$m'=m_i+\Delta m_i$ ($< m$) at time $t+\Delta t$,
and that each of these galaxies provides mass $m-m'$ for the background.
The number of the galaxies does not change during
this mass loss, unless they are completely disrupted.

Since we use discrete mass components,
a new mass $m'$ is generally not equal to any components $m_j$,
and therefore we have to somehow distribute the galaxies
of mass $m'$ among some components.
As Lee (1987) and Quinlan and Shapiro (1989)
did for stellar mergers in star clusters,
we distribute the galaxies by linear interpolation:
if $n$ galaxies of mass $m'$ are to be formed
and if  $m_j \le m' \le m_{j+1}$,
then $n(m_{j+1}-m')/(m_{j+1}-m_j)$ galaxies are added to the $m_j$ component
and $n(m'-m_j)/(m_{j+1}-m_j)$ galaxies are added to the $m_{j+1}$ component.
If $m' < m_1$, where $m_1$ is a minimum galaxy mass assumed,
the galaxies are distributed between $m_0=0$ (background) and $m_1$.
Further we assume that
the galaxies after mass stripping and the ejected background particles
have the same orbital energy and angular momentum per unit mass $(E,R)$
as the galaxies had before stripping.
The effect of energy dissipation is
considered separately, as described below.

By doing the above procedure over all of the galaxy components
at each grid point in $(E,R)$ space, 
we can calculate the change in the distribution function due to
mass stripping, i.e., the source term, $S_i$.
Note that this procedure guarantees conservation of the total mass and energy
of the cluster.

Merritt (1983) formulated mass stripping
as a first-order ``mass diffusion'' term in the FP equation
[see his equation (22)].
In his formulation he assumed a distribution function, $f(E,m,t)$,
which is continuous with respect to $m$ as well as $E$.
In integrating the FP equation numerically,
he used, of course, discrete mass grids.
On the other hand,
we have assumed discrete mass components from the beginning.
Although
Merritt's formulation and our formulation may look different
apparently,
they both consider
only the first-order term $\langle \Delta m \rangle$ and 
are essentially the same.

The effect of energy dissipation (cooling) is
included in the FP equation in the same way
as the effect of binary heating in star clusters (see, e.g., Takahashi 1997).
That is, the cooling term, which may be calculated from equation (\ref{eq:edis}),
is added to the first-order energy coefficient, $D_E$. 
In practice, the amount of energy dissipation during $\Delta t$ is
calculated so that it should be consistent with the above-described
way of distributing stripped galaxies among mass components.

\subsection{Time Scales}\label{sec:fpmodel5}

A commonly used definition of the two-body relaxation time is given by 
Spitzer (1987, equation [2-61]) as 
$t_{\rm r} \equiv (v_{\rm m}^2/3)/\langle (\Delta v_\parallel )^2
\rangle_{v=v_{\rm m}}$.
Here, a Maxwellian velocity distribution
of field particles is assumed and
$v_{\rm m}^2$ is the three-dimensional velocity dispersion;
$\langle (\Delta v_\parallel )^2 \rangle_{v=v_{\rm m}}$
is the mean square value of the change per unit time 
in the velocity component parallel to the
initial velocity of a test particle, evaluated at $v=v_{\rm m}$.
For the present cases, assuming that all galaxy components
follow Maxwellian velocity distributions 
with the same velocity dispersion, $V_{\rm cl}$,
we obtain
\begin{equation}
T_{\rm r} = \frac{0.065 V_{\rm cl}^3}{G^2 \ln \Lambda
                  \displaystyle{\sum_f m_f} \rho_f }
=\frac{0.065}{G^2 \ln \Lambda} \frac{V_{\rm cl}^3}
 {\langle m_f \rangle_\rho \rho_{\rm gal}} , \label{eq:tr}
\end{equation}
where $\rho_{\rm gal} = \sum \rho_f$ is the total density of the galaxies
and $\langle m_f \rangle_\rho = \sum m_f \rho_f/\rho_{\rm gal}$
is the {\it mass-weighted} mean galaxy mass.
Equation (\ref{eq:tr}) represents the time scale of the velocity diffusion
due to encounters between galaxies.
Similarly, we define the dynamical friction time
as $t_{\rm df} \equiv |v_{\rm m}/\langle \Delta v_\parallel \rangle_{v=v_{\rm m}} |$,
where $\langle \Delta v_\parallel \rangle_{v=v_{\rm m}}$ is the mean change per unit time
in the velocity component parallel to the initial velocity,
evaluated at $v=v_{\rm m}$.
On the same assumptions as used in deriving equation (\ref{eq:tr}),
we find that
the dynamical friction time for a galaxy of mass $m$ is given by
\begin{equation}
T_{\rm df} = \frac{0.13 V_{\rm cl}^3}{G^2 \ln \Lambda
                   \displaystyle{\sum_f (m+m_f)} \rho_f} 
           = \frac{0.13}{G^2 \ln \Lambda} \frac{V_{\rm cl}^3}
             {\langle m_f \rangle_\rho \rho_{\rm gal}+m \rho_{\rm tot}}, 
             \label{eq:tdf}
\end{equation}
where $\rho_{\rm tot}$ is the total density including
both the galaxy and halo components.
Unlike the relaxation time (\ref{eq:tr}),
the dynamical friction time (\ref{eq:tdf}) depends on the test-particle mass
and on the halo density.
From equations (\ref{eq:tr}) and (\ref{eq:tdf}), it follows that
\begin{equation}
\frac{T_{\rm r}}{T_{\rm df}} = 0.5 \left( 1 + \frac{m}{\langle m_f
\rangle_\rho} \frac{\rho_{\rm tot}}{\rho_{\rm gal}} \right) .
\end{equation}
In single-component clusters, $T_{\rm r}=T_{\rm df}$.
In multicomponent clusters, $T_{\rm df}$ can be much smaller than
$T_{\rm r}$ for massive galaxies; this holds also for light galaxies,
if $\rho_{\rm tot} \gg \rho_{\rm gal}$.

We define the tidal mass stripping time as $ t_{\rm ts} \equiv |m/(dm/dt)| $.
Using equation (\ref{eq:ml4}) with $V_f^2$ set equal to $2 V_{\rm cl}^2$,
we find
\begin{equation}
T_{\rm ts} 
= \frac{m_{\rm g0}^{2-(4-\eta_v)\zeta }}{C r_{\rm g0}^2 v_{\rm g0}^{\eta_v}}
\frac{m^{\eta_m-2+(4-\eta_v)\zeta}(\sqrt 2 V_{\rm cl})^{\eta_v-1}}
{\langle m_f^{\eta_m-1} \rangle_\rho \rho_{\rm gal}}. \label{eq:tts} 
\end{equation}
Using equations (\ref{eq:tdf}) and (\ref{eq:tts}), with $r_{\rm g0}=G
m_{\rm g0}/2v_{\rm g0}^2$, we obtain
\begin{equation}
\frac{T_{\rm ts}}{T_{\rm df}} =
\frac{31\ln \Lambda}{C}
2^{(\eta_v-1)/2} \left( \frac{m}{m_{\rm g0}} \right)^{(4-\eta_v)\zeta}
\left( \frac{v_{\rm g0}}{V_{\rm cl}} \right)^{4-\eta_v}
\frac{\langle m_f \rangle_\rho \rho_{\rm gal} + m \rho_{\rm tot}}
{m^{2-\eta_m} \langle m_f^{\eta_m-1} \rangle_\rho \rho_{\rm gal}} .
\end{equation}
For our standard set of parameters,
$\eta_m=2$, $\eta_v=2$, and $\zeta=1/4$ (see section \ref{sec:r1}),
it follows that
\begin{equation}
\frac{T_{\rm ts}}{T_{\rm df}} =
1.8 \ln \Lambda \left(\frac{25}{C}\right)
\left( \frac{m}{m_{\rm g0}} \right)^{1/2}
\left( \frac{v_{\rm g0}}{V_{\rm cl}} \right)^{2}
\left( 1+ \frac{m}{\langle m_f \rangle_\rho} \frac{\rho_{\rm
tot}}{\rho_{\rm gal}} \right) .  \label{eq:ttstdf}
\end{equation}
This indicates that $T_{\rm ts} \approx T_{\rm df}$ in our initial
models (see section \ref{sec:ic}).
As the fraction of the common halo increases,
tidal stripping becomes less important (Merritt 1983).

\section{Initial Conditions}\label{sec:ic}

We use the same initial cluster model as that in SFM99.
That is, the initial model is a Plummer model composed of $N_{\rm g}=128$
identical galaxies with no common halo.
Exceptions are models FPA1 and FPA2, which we describe later.
The initial galaxy model is also given by a Plummer model as in SFM99.
The ratio of the virial radius of each galaxy $r_{\rm vr}$
to the virial radius of the cluster $R_{\rm vr}$ is set equal to $1/20$.

Note that initial galaxy models affect FP models 
only through the mass-stripping rate, equation (\ref{eq:ml4}).
This equation includes two parameters of the internal structure of galaxies,
$r_{\rm g0}$ and $v_{\rm g0}$, but
these are related by the virial theorem (we take $r_{\rm g}=r_{\rm
vr}$) for a given galaxy's mass.
Hence, in FP models the structure of galaxies
is represented by only one parameter, e.g., $r_{\rm vr}$.
In reality, even if the virial radii are the same,
different galaxy models will give different stripping rates.
Such differences in galaxy models could be reflected in FP models only
by adjusting an undetermined numerical constant, $C$, in equation (\ref{eq:ml4}).

We use a system of units in which
$G=1$, $m_{\rm g}=1$, and $\varepsilon_{\rm g}=1/4$,
where $G$ is the gravitational constant (Heggie, Mathieu 1986);
hence $r_{\rm vr} = G m_{\rm g}^2/(4\varepsilon_{\rm g})=1$.
For the cluster, we have
$M_{\rm cl}=128$, $R_{\rm vr}=20$, and ${\cal E}_{\rm cl}=204.8$,
where $M_{\rm cl}$, $R_{\rm vr}$, and ${\cal E}_{\rm cl}$ are
the mass, virial radius, and binding energy of the cluster, respectively.
The mean velocity dispersion of stars in a galaxy
is $v_{\rm g}=1/ \sqrt 2$, and that of galaxies in the cluster
is $V_{\rm cl}=4/\sqrt 5$.
The cluster crossing time is
\begin{equation}
T_{\rm cr}=\frac{2R_{\rm vr}}{V_{\rm cl}} =10 \sqrt 5 \simeq 22,
\end{equation}
and the half-mass relaxation time (Spitzer 1987, equation [2-63]) is
\begin{equation}
T_{\rm rh}=0.138 \frac{N}{\ln \Lambda}
\left(\frac{R_{\rm h}^3}{GM_{\rm cl}}\right)^{1/2}
\simeq 19 ,
\end{equation}
where the cluster half-mass radius $R_{\rm h}=15.4$ and
the Coulomb logarithm $\ln \Lambda = \ln N = \ln 128$.
If we assume that $m_{\rm g}=10^{12} \MO$ and $r_{\rm vr}=30~{\rm
kpc}$, the cluster model has
$M_{\rm cl}=1.28\times 10^{14}\MO$, $R_{\rm vr}=0.6~{\rm Mpc}$,
$T_{\rm cr}=2.5~{\rm Gyr}$, and $T_{\rm rh}=2.1~{\rm Gyr}$.

Only models FPA1 and FPA2 have 
a common halo initially (see subsection~\ref{sec:r1.1});
the ratio of the common halo mass to the total mass, $M_{\rm h}/M_{\rm cl}$,
is $1/2$ and $3/4$, or if the mass is converted into the number of galaxies,
$N_{\rm g}=64$ and 32,
for FPA1 and FPA2, respectively.
Galaxy and common halo components are distributed in space so that
the ratio of their densities is independent of the radius.
The other conditions are the same as those of the other models.

\section{Simulations}

\subsection{Fokker--Planck Simulations}

Details of the numerical integration scheme of the FP equation have been
described by Takahashi (1995, 1997).
We used 301 energy, 51 angular momentum,
and 121 radial grid points.
We used $K=20$ mass components for galaxies.
These components are uniformly spaced in $m$ between 0 and 1;
$m_1=1/K$, $m_2=2/K$, \dots, $m_K=1$.
In addition, one component of $m_0=0$
was assigned to a background (common halo).
We confirmed that
these numbers of grid points and mass components
are large enough to obtain satisfactory convergence of the results.
The relative errors in the total mass were less than 0.1\%
and those in the total energy were less than 0.5\% in all runs.

We calculated a number of FP models, switching on and off mass
stripping and energy dissipation and changing the parameters
of the stripping rate.
These models are summarized in table~1.

\subsection{$N$-body Simulations}

We compared our FP models with the $N$-body models of SFM99 and SFM00
for the same initial conditions.
SFM99 performed a number of simulations for the same initial
theoretical model using different random realizations of the model
and varying the number of particles for representing each galaxy;
they confirmed that the results of those simulations were almost the same.
In section~\ref{sec:r1},
the model PP of SFM00, denoted by NB in this paper,
is compared with the FP models. 
The simulation was carried out using GRAPE-4 (Makino et al.~1997)
with the Barnes--Hut tree algorithm (Barnes, Hut 1986; Makino 1991;
Athanassoula et al.~1998).
Each initial galaxy model was represented by 2048 particles, and thus
the total number of particles was 262144.
For more details of the numerical integration and for a galaxy
identification scheme, see SFM99 and SFM00.

\section{Comparison between Fokker--Planck and $N$-body Models}\label{sec:r1}

\subsection{Models with Common Halos but without Mass Stripping}
\label{sec:r1.1}

First
we discuss the evolution of model clusters in which 
mass stripping from galaxies is not allowed, but
which initially have massive common halos,
in order to reveal how such multicomponent systems evolve
owing to two-body relaxation only.

Although collisional evolution of multicomponent stellar systems
has been investigated 
by many authors using FP models in the context of globular cluster
evolution (e.g., Inagaki, Wiyanto 1984; Cohn 1985; Chernoff, Weinberg
1990; Takahashi 1997 and references therein),
massless particle components were not considered in those studies.
On the other hand,
it is not very difficult to extrapolate the behavior of massless
particles
from the behavior of particles whose masses are much smaller than the mean
particle mass.
In that sense,
no essentially new results were found in our simulations, as discussed 
below. 
However, it is useful to see first
how ``pure'' collisional systems evolve
before considering more complicated systems with mass stripping.
Yepes et al.\ (1991) and Yepes and Dom\'\i nguez-Tenreiro (1992) also
studied pure collisional evolution of galaxy clusters including massless
particles, but using the moment equations of the FP equation.


Figure 1a shows the evolution of density profiles for model FPA1;
the solid lines and dotted lines
represent the galaxy and common halo components, respectively.
Figure 1b shows the corresponding profiles of the
logarithmic gradient defined by
\begin{equation}
\alpha_i \equiv -\frac{d\ln \rho_i}{d \ln r} . \label{eq:alpha}
\end{equation}
Figures 1c and 1d are for model FPA2.

We have assumed that the number of galaxies, $N_{\rm g}$, is
64 and 32 for FPA1 and FPA2, respectively.
However, it should be noted that
$N_{\rm g}$ is only concerned with time scaling;
if time is measured in units of the relaxation time,
the evolution of these models does not depend on the number of
galaxies, because only the two-body relaxation process is taken into account
in them.
We set the Coulomb logarithm equal to
$\ln (M_{\rm cl}/m_{\rm g})=\ln 128$ for time scaling for both models.

As is well known, energy redistribution among particles due
to two-body relaxation eventually leads the cluster core to gravothermal
core collapse (Lynden-Bell, Wood 1968; Lynden-Bell,
Eggleton 1980); after the onset of core collapse
the central density continues to increase and the core radius
continues to decrease, unless other physical mechanisms intervene.
Late core collapse proceeds self-similarly with the development of
the power-law density cusp, $\rho \propto r^{-\alpha}$.
Cohn (1980) found $\alpha=2.23$ for single-component clusters.
In multicomponent clusters, the most massive component eventually
dominates the central part, i.e., strong mass segregation occurs,
owing to a tendency toward equipartition of energy.
Thereafter, the most massive component collapses as in single-component clusters
and lighter components have shallower density cusps
(e.g., Inagaki, Wiyanto 1984; Chernoff, Weinberg 1990).

The development of the density cusps of the galaxy and common-halo components
through gravothermal core collapse
is clearly shown in figure 1.
Power-law cusps appear for $r \ltsim 1 = R_{\rm vr}/20$.
The slopes of the cusps hardly depend on the difference in
the initial conditions of FPA1 and FPA2;
the slope of the galaxy cusp approaches $\alpha=2.23$
and the slope of the common-halo cusp is much shallower,
$\alpha \sim 0.5 $,
as expected.

Cohn (1985; see also Chernoff, Weinberg 1990) found that
in deep collapse phases of multimass clusters,
the power-law indices $\alpha_i$ of the density profiles of different
components are approximately related by
\begin{equation}
\alpha_i
= 0.23  \left( 8.2\frac{m_i}{m_{\rm u}} +\frac{3}{2} \right) ,
\label{eq:alphai}
\end{equation}
where $m_{\rm u}$ is the particle mass of the most massive component,
which dominates the central potential well.
This equation is obtained on the assumption that
the distribution function of component $i$ is given by
$f_i \propto (-E)^{p_i}$ with $p_i/p_{\rm u}=m_i/m_{\rm u}$.
This relation for $p_{\rm i}$ was originally found by Bahcall and Wolf (1977)
for a steady state solution of the stellar distribution around a
central massive black hole.
In deriving equation (\ref{eq:alphai}) the results of numerical simulation
(Cohn 1980), $\alpha_{\rm u}=2.23$ and $p_{\rm u}=8.2$, are also used.

Equation (\ref{eq:alphai}) gives $\alpha_i \to 0.345$ as $m_i \to 0$.
Figures 1b and 1d show that the actual slope of the common halo cusp
is slightly larger than this predicted value.
Such deviation from equation (\ref{eq:alphai}) at small masses
was also reported by Chernoff and Weinberg (1990).

For models FPA1 and FPA2, we followed the cluster evolution
up to a rather deep collapse phase in order to see self-similar
development of the density cusps.
However, we should keep in mind that
it is very unlikely that actual galaxy clusters can reach
such deep collapse, because galaxies have appreciable finite sizes
compared to the cluster sizes.
For our model cluster,
the sizes of the galaxies are $\sim 1$,
and thus the core radius of the cluster cannot be less than $\sim 1$.

\subsection{Models with Mass Stripping but without Energy Dissipation}
\label{sec:r1.2}


In this section we show the results of FP models in which mass stripping
is included but energy dissipation is neglected,
and compare these results with $N$-body results.

Figure~2a shows the fraction of the common halo mass to the total mass
as a function of time for models FPB1 and NB.
Figure~2b shows the evolution of the central density of
the galaxies and that of the common halo, $\rho_{\rm g}(0)$ and $\rho_{\rm h}(0)$.
For the $N$-body model the galaxy density is not plotted,
because it changes very noisily with time owing to
the small number of galaxies (see figure 6).
The center of the cluster of the $N$-body model was chosen to be
the density center (Casertano, Hut 1985) of the common halo particles.

Concerning the growth of the common halo mass, $M_{\rm h}$,
the agreement between FPB1 and NB is pretty good,
though the growth speed slows down somewhat earlier in FPB1 than in NB.
Actually, 
the value of the numerical constant, $C=25$,
in equation (\ref{eq:ml4}) for FPB1 
was chosen to achieve such a good agreement.
The initial growth rate of $M_{\rm h}$ in the FP model is
determined by only the stripping rate
for given initial conditions.
This means that we can always adjust $C$ to obtain a desirable
initial growth rate.
Once its value is chosen, however, we have no freedom to adjust later
growth of the common halo, which depends on evolving cluster properties.
Therefore,
the fine tuning of $C$ is not responsible for
the overall good agreement observed in figure 2a.

In principle,
the value of $C$ can be determined from the data of $N$-body
experiments of galaxy--galaxy encounters such as given by FM.
However, its exact value depends on the detailed structure of the galaxies,
which changes with time under the cluster environment.
From equations (\ref{eq:ml2}) and (\ref{eq:ml3}), we find,
for $\eta_v=2$,
$C'=C(8\pi/3)^{-1/2}/ \ln(R_{\rm vr}/r_{\rm vr})= 0.115 C$
if $\eta_r=2$, and
$C'=C(8\pi/3)^{-1/2}/ (1/r_{\rm vr}-1/R_{\rm vr})= 0.364 C$
if $\eta_r=3$.
Hence $C=25$ corresponds to $C'=2.9$ and 9.1 for $\eta_r=2$ and 3,
respectively.
Equation (\ref{eq:ml0}) with these values of $C'$ is consistent,
in order of magnitude, with the data of FM
(see their figures 7, 8, 11, and 12).

Models FPB1 and NB are in rather good agreement
in the evolution of $M_{\rm h}$,
but show a clear difference in $\rho_{\rm h}(0)$ evolution:
the central density increases much more slowly in FPB1 than in NB.
We will see below that the inclusion
of energy dissipation, which makes the treatment of tidal stripping
in FP models fully self-consistent,
removes this difference.

The initial quick increase in $\rho_{\rm h}(0)$
and the initial decrease in $\rho_{\rm g}(0)$ are
simply due to the fact that mass stripping from the galaxies suddenly starts
at $t=0$.
After a few cluster crossing times, the central galaxy density
begins to increase.
Mass stripping works only to decrease the galaxy density
by removing part of galaxy mass to the common halo.
Therefore, the increase in the galaxy density should be due to two-body relaxation.
Since the processes of core collapse and mass stripping occur simultaneously,
the central density of the galaxies and that of the common halo 
stay comparable for a long time,
while the central density is soon dominated by galaxies
in models FPA1 and FPA2 where there is no mass stripping.
Figure~2b shows that
the galaxy density finally
exceeds the common halo density at the center even with mass stripping.
We discuss in more detail the process of 
core collapse with mass stripping in section \ref{sec:r2}.

Here, we mention the choice of the value of the Coulomb logarithm.
As noted above, the central density increase
in model FPB1 is significantly slower than that in model NB.
We found that rough agreement in the central density evolution was
obtained
when we artificially increased
the two-body relaxation rate by about a factor of two without
changing the mass stripping rate.
Increasing the relaxation rate by a factor of two is equivalent to
doubling the Coulomb logarithm, $\ln \Lambda$, i.e., replacing $\ln 128$ with
$2\ln 128 = \ln 128^2$ in the present case.
There is some uncertainty for the choice of the exact value of $\Lambda$,
but it is very unlikely that such a large value as $\Lambda =128^2$
is justified.

\subsection{Models with Mass Stripping and Energy Dissipation}
\label{sec:r1.3}


We now take account of the effects of both mass stripping and
energy dissipation in FP models (sequence FPC in table~1).
The evolution of model FPC1 (our standard model) is shown
in figure 3.
The growth of $M_{\rm h}$
in FPC1 is similar to that in FPB1,
although the former shows a slightly larger growth rate at late epochs.
The difference between FPB1 and FPC1 is clear in the evolution of
the central density,
which increases more rapidly in FPC1 than in FPB1.
We now see good agreement between models FPC1 and NB
in the central density 
evolution as well as in the evolution of the common halo mass.
It should be recalled that energy dissipation is included
consistently with mass stripping, without introducing any additional free
parameters.


In the following,
we investigate how sensitively the behavior of FP models
depends on the parameters of the mass stripping rate
($C$, $\eta_v$, $\eta_m$, and $\zeta$).
The probable ranges of these parameters
are limited by the results of $N$-body experiments
and theoretical considerations (see section \ref{sec:fpmodel2}).

In figure 4 we compare models FPC1, FPC2, and FPC3, which differ only
in the value of $C$ ($C=$ 25, 15, 35). 
Larger $C$ gives a higher stripping rate, and consequently
faster growth in the mass of the common halo.
At early epochs the central density of the common halo also increases
more quickly for larger $C$,
but at late epochs it increases more slowly because
faster mass stripping prevents the density increase more severely.
We conclude that the value of $C=25$ gives the best agreement between the FP and
$N$-body results.


Figure 5 compares models FPC1, FPC4, FPC5, and FPC6, which differ
in the value of $\eta_v$, $\eta_m$, or $\zeta$ (see table 1).
For model FPC4 it is necessary to adjust the value of $C$ so that
the initial growth rate of the common halo becomes similar to those
of the other models.
These models are not very different from one another.
Therefore, we conclude that the results of FP simulations
do not very sensitively depend on the mass-stripping formula as long as
the parameters are in reasonable ranges.

One noticeable difference is seen in the evolution of the central
density of model FPC5 ($\eta_m=1$); at late epochs
the central density increases
more slowly compared with the other models.
We find that this is because massive galaxies are much more depleted
in model FPC5.
In all models,
the mean galaxy mass $\langle m_f \rangle_\rho$ in the central regions 
first decreases quickly
and then stays roughly constant as the galaxy density increases.
At late epochs, $\langle m_f \rangle_\rho \sim 0.2$ at the center
for model FPC5
and $\sim 0.4$ for the others. This implies that the collapse rate
of FPC5 is smaller by about a factor of two.
Equation (\ref{eq:tts}) tells us why such a difference occurs;
tidal stripping time
$T_{\rm ts}$ is proportional to $1/ \langle m_f \rangle_\rho$ for $\eta_m=2$,
but $T_{\rm ts}$ does not depend on $\langle m_f \rangle_\rho$ for
$\eta_m=1$.
Therefore, while 
$T_{\rm ts}$ increases as $\langle m_f \rangle_\rho$ decreases
in the $\eta_m=2$ models,
it stays almost constant (actually somewhat decreases) in model FPC5.
Consequently, $\langle m_f \rangle_\rho$ decreases to a much lower
value in FPC5.
A comparison with the $N$-body model in figure 5b supports $\eta_m=2$
rather than $\eta_m=1$.



Figure 6 shows the mass-density profiles 
of models NB (left) and FPC1 (right) at selected epochs.
The density profiles for the galaxies (middle) and the common halo (bottom)
are plotted separately as well as the total density (top).
For model NB, the density profiles of the galaxies are rather noisy 
because of the small number of galaxies.
We should note that
a comparison of the profiles of the $N$-body and FP models
on a length scale of $\ltsim 1$ is, at least for the galaxies,
meaningless, since the sizes of the galaxies, $r_{\rm g}$, is $\sim 1$. 
Thus, for $r \ltsim 1$ in the $N$-body model,
we see the internal structure of a galaxy
if any galaxy is there, and a void if not.
At late epochs, the central region of the cluster is usually
occupied by a few massive galaxies
that have sunk there owing to dynamical friction.
For $r \gtsim 1$, models NB and FPC1 show
very similar evolution of the density profiles, apart from
fluctuations in the $N$-body model.

The similarity between the two models is more clearly shown
in figure 7, which directly compares their density profiles
at $t=8.9\,T_{\rm cr}$.
In particular, almost perfect agreement is seen
in the density profile of the common halo.

The agreement between the FP and $N$-body results
shown above
is enough to convince us that 
our FP models include all essential physical
processes concerned and can describe the cluster evolution rather accurately.


In figure 8 we show the logarithmic density gradient, $\alpha$
[see equation (\ref{eq:alpha})],
as a function of the radius for model FPC1 at $t=8.9\,T_{\rm cr}$, together with the density profiles
of FPC1 and NB again.
For $1 \ltsim r \ltsim 5$, the value of $\alpha$
for the total density
varies gently with the radius from $\sim 0.5$ to $\sim 1.5$
with a median value of $\sim 1$.
This implies that the total density profile in this region
may be reasonably approximated by a power law, $r^{-1}$, as SFM99 claimed.

The common halo component dominates the density for
$1 \ltsim r \ltsim 20$.
For $r \ltsim 1$, the density of the galaxy component overwhelms
that of the halo component in the $N$-body model,
while they are comparable in model FPC1.
As noted above, this difference arises because 
in the $N$-body model a massive galaxy stays around the cluster center
at late epochs and
its internal structure is observed on this length scale.
In fact, the density of the particles in each galaxy
is much larger than the density of the common halo particles
for $r \ltsim 1$ in NB.
This extra density increase in the $N$-body model
strengthens our impression that the total density profile is given by
a power law.

It is not clear from figure 8 whether the central density
distribution is really described by any single power law,
since the central density has not yet increased sufficiently from the
initial value.
In principle, FP calculations can be (formally) continued further,
if we neglect the fact that the galaxies have finite sizes.
However, we could actually not continue FP calculations
to very advanced stages of
core collapse for model FPC1 and other similar models,
unlike the case for the models without stripping,
owing to numerical instability.
This difficulty of FP calculations is linked to the divergence of
the distribution function at the center caused by continuous decrease
in the central velocity dispersion; see section \ref{sec:r2}.



In order to see the mass dependence of the galaxy distributions,
we divided the galaxies into four mass groups,
$m \in (0,0.25]$, $(0.25,0.50]$, $(0.50,0.75]$, and  $(0.75,1]$,
and plotted the density profile for each group
for model FPC1 at $t=8.9\,T_{\rm cr}$ in figure 9.
It is apparent that heavy galaxies are strongly depleted in the central regions,
owing to tidal stripping, and have a very flat density profile.
If there were no stripping,
the heavy galaxies would dominate the central regions
as the result of two-body relaxation.

The development of ``inverse mass segregation'' is clearly shown in
figure 10, which plots the mass-weighted mean galaxy mass,
$\langle m \rangle_\rho$, as a function of the radius for model FPC1.
The mean mass decreases toward the center.
This indicates that mass stripping is more efficient
than dynamical friction for heavy galaxies.
Only at very late epochs ($t \gtsim 8.9\,T_{\rm cr}$),
when the central density of the galaxies has increased sufficiently high,
is a slight increase in the mean mass seen at the center. 
Similar inverse mass segregation was already found by
Funato et al.\ (1993) and SFM99 in $N$-body simulations
and by Merritt (1984) in FP simulations.

The density profiles of models FPC1--C6 are similar,
if they are compared at epochs with the same central density.
In particular, the halo density profiles are almost indistinguishable.
However, we show in the next section that the density cusp gradient
does depend on the stripping rate. 
It is obvious that the gradient, $\alpha$, should be close to 2.2
when mass stripping is almost negligible compared to two-body relaxation.
On the other hand, if stripping is infinitely fast, all galaxies
will dissolve instantaneously and no further evolution of the cluster
can occur based on our present assumptions.
The stripping rates for models FPC1--C6 are not very different 
(see figures 4 and 5), and hence the difference in their density 
profiles is insignificant.

\section{Cool Core Collapse and Shallow Density Cusps}\label{sec:r2}

In subsection \ref{sec:r1.3} we showed that a shallow density cusp,
approximated by $\rho \propto r^{-1}$,
develops in the central regions of a galaxy cluster.
Now let us assume that a collapsing core leaves behind
a cusp that exactly follows a power-law, $\rho \propto r^{-\alpha}$. 
This implies that the central density
grows as $\rho_{\rm c}(t) \propto r_{\rm c}^{-\alpha}(t)$,
as the core radius $r_{\rm c}(t)$ decreases with time. 
Similarly, a dimensional analysis implies that
the central velocity dispersion changes as $v_{\rm c}^2 (t) \propto
G\rho_{\rm c} r_{\rm c}^3/r_{\rm c} \propto r_{\rm c}^{2-\alpha}(t)$
(but this is valid only for $\alpha > 1$ as we discuss later). 
This relation predicts that
the central velocity dispersion increases (i.e. the core gets hotter)
as the core shrinks if $\alpha >2$,
and that it decreases (i.e. the core gets cooler) if $\alpha < 2$.
For core collapse in star clusters, detailed numerical calculations
(Cohn 1980; Lynden-Bell, Eggleton 1980) show that $\alpha =2.2 > 2$,
and that actually both the central density and velocity
dispersion increase as the core shrinks --- ``hot collapse'' occurs.
For core collapse in galaxy clusters, our calculations indicate $\alpha < 2$.
Therefore, we expect ``cool collapse'' in this case.

Lynden-Bell and Eggleton (1980) proved that $\alpha$ should be in the
range $2 < \alpha < 2.5$ for the existence of any self-similar collapse
solution in equal-mass clusters. The condition $\alpha > 2$ comes from
the requirement that temperature must decrease outward so that
the core can shrink by losing heat to the halo.
In multimass clusters, however,
heat is also transferred among different mass components.
Therefore, it can happen that the core of heavy particles loses energy
to light particles and shrinks even if $\alpha < 2$.

Let us see in more detail the case of a shallow density cusp,
\begin{equation}
\rho(r) \propto r^{-\alpha}  \quad\quad (0< \alpha <2).  \label{eq:rho.sc}
\end{equation}
This produces a potential, $\phi(r)$, that is finite at the center
and satisfies
\begin{equation}
\phi (r) -  \phi (0)  \propto r^{2-\alpha}. \label{eq:phi.sc}
\end{equation}
If we assume that the velocity distribution is isotropic
(this is not a bad assumption near the center),
we find that the density distribution (\ref{eq:rho.sc})
corresponds to the distribution function
\begin{equation}
f(E) \propto [E-\phi(0)]^{-(6-\alpha)/2(2-\alpha)}.  \label{eq:df.sc}
\end{equation}
Thus, $f$ diverges as $E \to \phi(0)$.
This divergence causes difficulty in numerical integration of the FP
equation, and therefore
we had to stop our calculations at relatively
early (low central density) collapse phases for the models with mass stripping.
The depth of the central potential well becomes shallower 
as $\alpha$ decreases (Tremaine et al. 1994).

As is argued by Tremaine et al.\ (1994), the dependence of the velocity
dispersion, $v_{\rm m}^2$, on the radius for the density cusp represented by equation
(\ref{eq:rho.sc}) changes at $\alpha=1$.
We find 
\begin{equation}
v_{\rm m}^2 (r) \propto r^{2-\alpha},  \quad\quad \mbox{for $\alpha > 1$} ,
\label{eq:v2.sc.1}
\end{equation}
which is the equation expected from a dimensional analysis.
For $\alpha<1$, the behavior of the velocity distribution
is more subtle, since
it is dominated by high-energy parts of the distribution function,
which must deviate from equation (\ref{eq:df.sc}) if the total
mass is to be finite.
If we consider any finite-mass model,
we find 
\begin{equation}
v_{\rm m}^2 (r) \propto r^{\alpha}  \quad\quad \mbox{for $\alpha <1$} 
\label{eq:v2.sc.2}
\end{equation}
(see also Dehnen 1993).
For any $\alpha < 2$, $v_{\rm m}^2 \to 0$ as $r \to 0$,
and hence cool core collapse is expected.


In figure 11 we plot the evolution of the central densities and the
central velocity dispersions of the galaxy and common halo components
for models FPA1 and FPC1.
Here the velocity dispersion of the galaxies is the mass-weighted
average over all the galaxies.
The central density increases with time in both models,
but the behavior of the velocity dispersion is very different
between FPA1 and FPC1.
As we expected, the central velocity dispersion of the galaxies
does decrease as core collapse proceeds in model FPC1,
while it increases in model FPA1.
In model FPC1, even the central velocity dispersion of the common halo
finally decreases when core collapse accelerates.
It is considered that this is because cool halo particles
are being supplied from cool central galaxies.
(Remember we assumed that the particles stripped from a galaxy
initially have the same energy and angular momentum per unit mass
as those of its parent galaxy.)

We now consider why the velocity dispersion of the galaxies continue
to decrease when mass stripping is included.
Spitzer (1969) showed that equipartition among different mass
components cannot be achieved
if self-gravity of heavy components is strong.
It is instructive to recall here his simple model.
Consider a spherical system composed of particles of two different masses,
$m_1$ and $m_2$ ($m_1<m_2$).
We assume that the total mass of the light component is
much larger than that of the heavy component, $M_1 \gg M_2$,
and that the heavy component is concentrated at the center.
Then, the virial theorem for component 2 can be written as
\begin{equation}
v_2^2 = a \frac{GM_2}{R_2} + b \frac{GM_1 R_2^2}{R_1^3},
\end{equation}
where $v_2$ is the mean velocity dispersion for component 2,
$R_1$ and $R_2$ are the median radii of components 1 and 2, respectively,
and $a$ and $b$ are numerical constants.
{\it For fixed $M_1$, $M_2$, and $R_1$,} the right side has a minimum
value of
\begin{equation}
v_{\rm 2,min}^2 = 2(a^2b)^{1/3} \frac{G (M_1M_2^2)^{1/3}}{R_1}
\end{equation}
at $R_2= R_{\rm 2,crit}\equiv (aM_2/bM_1)^{1/3} R_1$.
The minimum of $v_2^2$ exists because
the heavy particles form a self-gravitating system by themselves
when $R_2 \ll R_{\rm 2,crit}$, but behave as a non-self-gravitating system
when $R_2 \gg R_{\rm 2,crit}$.
Therefore, after the concentration of the heavy component becomes
sufficiently high, its velocity dispersion increases as the heavy component 
loses energy and contracts.
However, if we now allow $M_2$ to change with $M_1+M_2$ fixed, 
we find that $v_{\rm 2,min}^2 \to 0$ and $R_{\rm 2,crit} \to 0$ as $M_2 \to 0$.
This implies that
the velocity dispersion of the heavy component
can continue to decrease to zero as the core shrinks,
if mass stripping proceeds at an appropriate speed.
Mass stripping prevents the heavy component from forming
an independent self-gravitating system.



We may expect that
core collapse changes its nature from cool collapse to hot collapse
as tidal stripping becomes less and less important
compared with two-body relaxation.
To confirm this, we ran models FPC7, C8, and C9,
which are the same as model FPC1 but
have $C=$ 2.5, 0.25, and 0.025, respectively.
In principle, we should change
the initial models of galaxies and clusters rather than $C$, which is
considered to be roughly constant, in order to see the evolution
of clusters with different stripping importance.
From equation (\ref{eq:ttstdf}) we find that stripping becomes less
important as $v_{\rm g0}/V_{\rm cl}$ increases
with the other parameters fixed.
Thus, very low stripping rates imply the condition $v_{\rm g0} \gg V_{\rm cl}$.
(cf. In star clusters the escape velocity from a star is much larger
than the velocity dispersion of stars.)
However, we should not directly apply the equations for tidal stripping
presented in subsections 2.2 and 2.3 to such cases,
since those equations are considered to be valid only for $V_{\rm cl}
\gtsim v_{\rm g0}$.
We therefore choose to change $C$ just to conveniently
control the stripping efficiency in our theoretical experiments.

Figure 12 shows the growth of the common halo mass for models FPC1
and FPC7--C9.
It is interesting that the difference in the growth rates at the late times 
is within an order of magnitude among these models
($[T_{\rm cr}/M_{\rm cl}][dM_{\rm h}/dt] \sim$0.03, 0.08, 0.05, and 0.01
for FPC1, C7, C8, and C9, respectively),
while $C$ varies over three orders of magnitude.
This may be explained by
a kind of self-regulation mechanism:
if $C$ is large, the galaxies quickly become more compact through
initial violent stripping, and then the stripping speed slows down;
if $C$ is small, the central density can increase to very high values,
and this finally increases the stripping rate at the central regions.

Figure 13 shows the evolution of the central density and velocity
dispersion of the galaxy component for models FPC1 and FPC7--C9.
The transition from cool collapse to hot collapse occurs at $C \sim
0.25$. In model FPC8 the central velocity dispersion increases 
until the central density increases by about a factor of $10^3$,
but decreases after that.
Equation (\ref{eq:ttstdf}) tells us that in the present models
the stripping time eventually
becomes shorter than the relaxation time as $V_{\rm cl}$ increases
toward infinity.
Thus, we expect that
for any small value of $C$, tidal stripping will become effective
and the velocity dispersion will begin to decrease when the velocity
dispersion has increased to a critical value,
which is inversely proportional to $C$.
The above results show that,
unless the stripping rate is decreased to a very low value
compared to the standard value, cool collapse occurs.
Therefore, under the conditions of real galaxy clusters
($V_{\rm cl} \gg v_{\rm g0}$), cool collapse is expected to occur.



Figures 14 shows the profiles of the density and the logarithmic density
gradient $\alpha$ for model FPC7 ($C=2.5$), and figure 15 shows those for
model FPC9 ($C=0.025$).
In model FPC7 a shallow density cusp of $\alpha < 2$ develops,
but this cusp is steeper than the cusp in model FPC1 (see figure
8).  In model FPC9 a deep cusp of $\alpha > 2$ develops, and $\alpha$
approaches 2.2 as the core collapse proceeds.
Unlike in no-stripping models FPA, the central density of the halo component
also significantly rises in model FPC9 due to continuous production
of new halo particles.


Figures 16a and b show 
the evolution of velocity dispersion profiles for models FPC7 and
FPC9, respectively.
In these figures
qualitative difference between models FPC7 and FPC9 is clearly seen.
In model FPC7 an inverse velocity dispersion gradient develops in the
central regions.

\section{Summary and Discussion}\label{sec:sum}

In this study,
we investigated the dynamical evolution of galaxy clusters
after virialization using FP models and compared these models
with the $N$-body models of SFM99 and SFM00.
Our FP models included the effects of gravitational two-body
encounters between galaxies
and between galaxies and common halo particles which are much
lighter than the galaxies.
Furthermore, tidal stripping of mass from the
galaxies to the common halo and accompanying dissipation
of the orbital kinetic energies of the galaxies were included
using revised cross-section formulae.

We found that
the results of the FP models agree very well with those of the $N$-body models
regarding the growth of the common halo mass and the evolution of the
cluster density profiles.
This indicates that all of the important physical processes that occur 
in the $N$-body models are properly included in the FP models.
Although there is some ambiguity in the cross sections
of mass stripping and energy dissipation, we found that
the results of the FP simulations do not sensitively depend
on the details of the cross-section formulae.

In the evolution of galaxy clusters, tidal mass stripping
from galaxies is very important.
Under the initial conditions that SFM99 and SFM00 employed,
where the total cluster mass is initially attached to galaxies,
more than half of the total mass turns into the common halo during
a first few crossing times.
Heavy galaxies sink toward the cluster center owing to dynamical friction.
However, a high density of galaxies results in strong stripping,
and thus it is usually seen in our models that
the mean mass of galaxies decreases toward the center,
which is inverse mass segregation.
Dissipation of the orbital kinetic energies of galaxies due to tidal stripping
also plays an important role in accelerating
the sinking of the galaxies toward the cluster
center, if the velocity dispersion of the stars in the galaxies
is not very small compared to that of the galaxies in the cluster.

As the central density increases, a cusp profile, approximated by a power law,
$\rho \propto r^{-\alpha}$, develops at the core region.
In the model clusters of SFM99 and SFM00, $\alpha$ for the total density
is $\sim $1. This density distribution consists of
the galaxy component with larger $\alpha$ (still $<2$)
and the common halo component with smaller $\alpha$.
The development of the density cusp is
a consequence of the collisional evolution of galaxy clusters.

The slope of the density cusp, $\alpha$, depends on the ratio of
the stripping rate to the relaxation rate.
When the effect of stripping is negligible compared to relaxation,
``hot core collapse'', which is
usual gravothermal collapse, occurs: i.e.,
as the core shrinks both
the central density and velocity dispersion
increase, and a deep cusp with $\alpha \sim 2.2 > 2$ develops.
When stripping is important, ``cool core collapse'' occurs: 
while the central density increases,
the central velocity dispersion decreases, and consequently
a shallow cusp ($\alpha < 2$) develops.
Faster tidal stripping results in shallower cusps.
Under the conditions of real galaxy clusters,
tidal stripping is rather effective and cool collapse is expected to occur.

One might think that this conclusion is in contradiction with
the results of SFM00, who showed using $N$-body simulations
that the profile with $\alpha \sim
1.2$ (1--1.5) develops regardless of the initial models of 
galaxies and clusters.
Actually, their results are consistent with the results of this paper.
SFM00 varied the initial models of galaxies and clusters
adopting Plummer, King (King 1966), and Hernquist models (Hernquist 1990),
but did not vary the number of galaxies and the ratio of the galaxy
virial radius to the cluster virial radius.
As a result, the growth rates of the common halo mass, or the mean
stripping rates are similar among all of their runs.
Our FP simulations show that while $\alpha$ depends on the stripping
rate, it does so only weakly.
Therefore, it is no wonder that
no significant difference in $\alpha$ has been detected
among the $N$-body models of SFM00.

Hot core collapse terminates with the formation of hard binaries.
Because of subsequent energy release from these binaries,
the core starts to expand (see, e.g., Spitzer 1987).
What is the end state of cool collapse?
Through cool collapse, low-energy galaxies accumulate around
the cluster center. They finally merge and form a massive
central galaxy. The central galaxy will further grow
by eating galaxies falling to the center in succession, and this
process will continue until all galaxies, except for the central one,
disappear. Thus, we will see at last one giant (cD ?) galaxy with
an enormous halo.
In the present study we did not include the merging process in FP models,
and hence we could not follow very late stages of the cluster evolution
with FP models.
Future $N$-body calculations as well as improved FP models
will reveal the consequences of cool collapse in more detail.

\par
\vspace{1pc}\par
This work was supported in part by the Research for the
Future Program of Japan Society for the Promotion of Science
(JSPS-RFTP97P01102).



\begin{table}
\caption{Fokker--Planck models.}
\begin{center}
\begin{tabular}{cccccl}
\hline
\hline
Model       &  $\eta_v$ & $\eta_m$ & $\zeta$ & $C$ & Note \\
\hline
FPA1         & $\cdots$ & $\cdots$  & $\cdots$ & 0  & 
                              $M_{\rm h}/M_{\rm cl}=1/2$ \\
FPA2         & $\cdots$ & $\cdots$ & $\cdots$  & 0  & 
                              $M_{\rm h}/M_{\rm cl}=3/4$ \\
FPB1         & 2   & 2   & 1/4 & 25  & No dissipation\\
FPC1         & 2   & 2   & 1/4 & 25  & Standard model\\
FPC2         & 2   & 2   & 1/4 & 15  & \\
FPC3         & 2   & 2   & 1/4 & 35  & \\
FPC4         & 3   & 2   & 1/4 & 100 & \\
FPC5         & 2   & 1   & 1/4 & 25 & \\
FPC6         & 2   & 2   & 1/3 & 25 & \\
FPC7         & 2   & 2   & 1/4 & 2.5  & \\
FPC8         & 2   & 2   & 1/4 & 0.25  & \\
FPC9         & 2   & 2   & 1/4 & 0.025  & \\
\hline
\end{tabular}
\end{center}
\end{table}

\clearpage


\begin{figure}
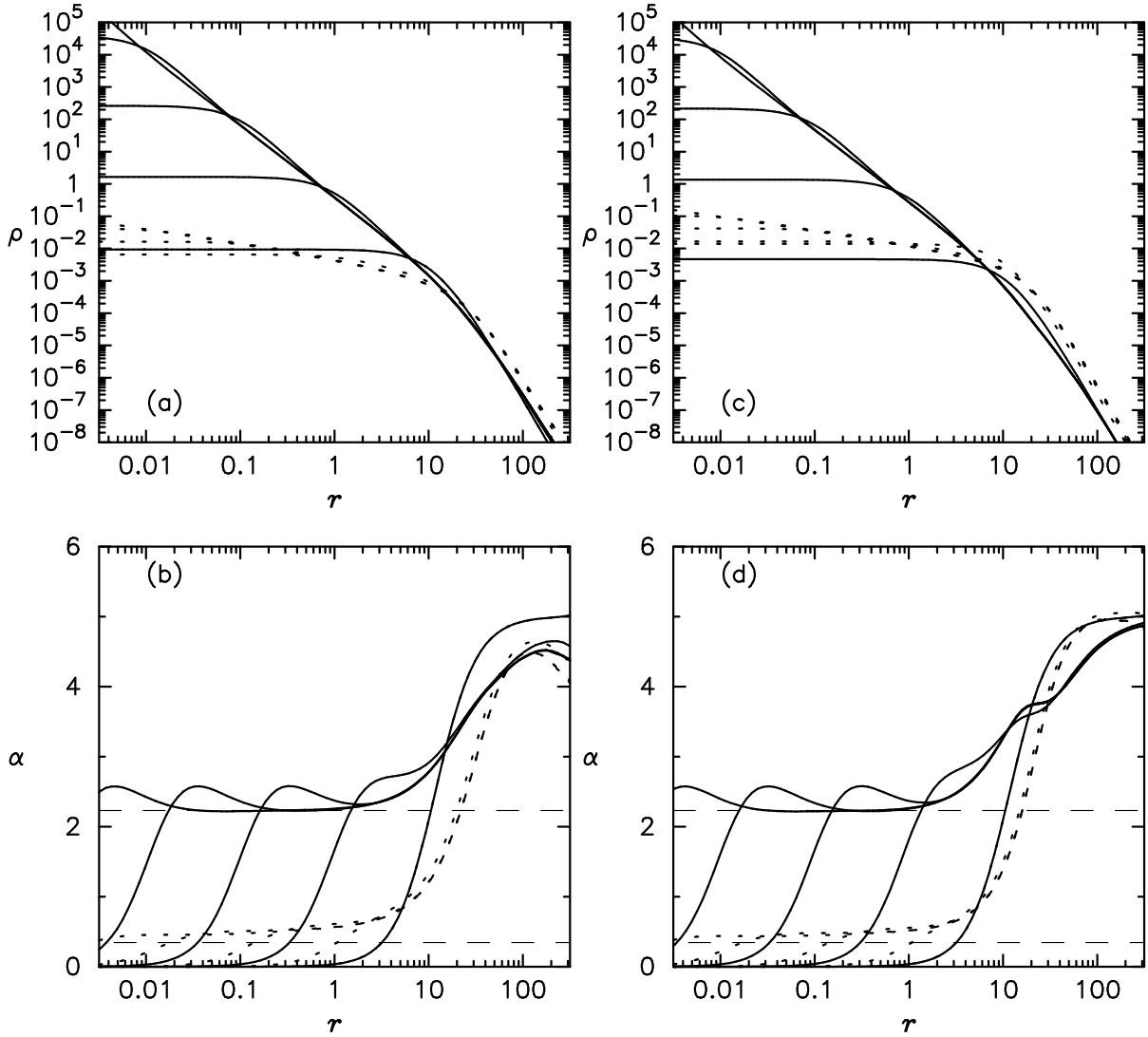

\begin{center}
\FigureFile(160mm,){fig1.ps}
\end{center}
\caption{(a) Evolution of the density profiles of the galaxy component
(solid) and common halo component (dotted)
for model FPA1.
Plotted are the profiles at $t/T_{\rm cr}=$ 0, 6.7, 7.99, 8.0209,
8.0214.
The central densities of both components increase with time,
except that the central density of the halo component decreases initially.
(b) The logarithmic density gradient $\alpha \equiv -d\ln\rho/d\ln r$ 
for the density profiles shown in (a).
The two horizontal dashed lines are $\alpha=2.23$ and $\alpha=0.345$
(see text).
(c) Same as (a), but for model FPA2.
Plotted are the profiles at $t/T_{\rm cr}=$ 0, 4.5, 5.42, 5.4454,
5.4457.
The central densities of both components increase with time.
(d) Same as (b), but for model FPA2.}
\end{figure}

\clearpage

\twocolumn

\begin{figure}
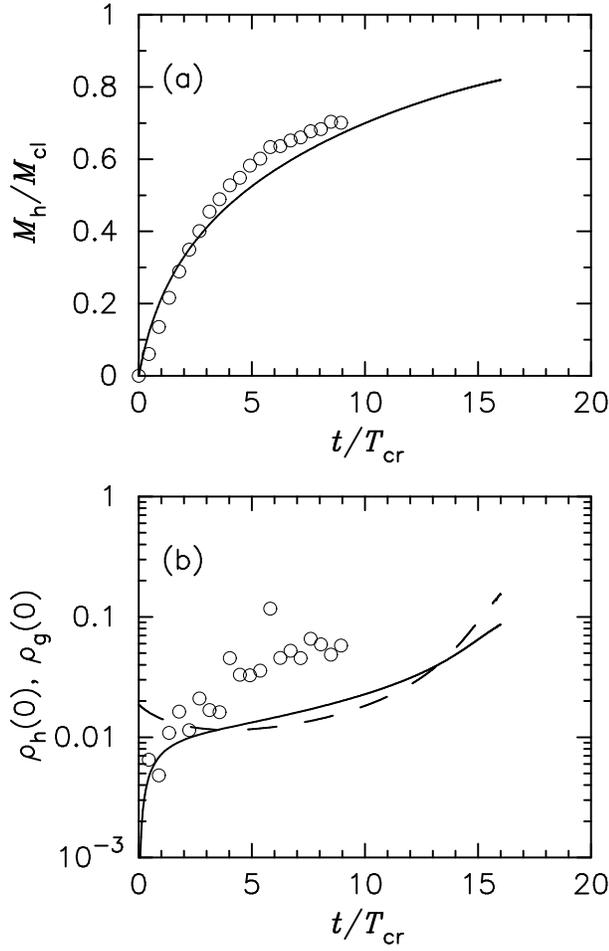

\begin{center}
\FigureFile(80mm,){fig2.ps}
\end{center}
\caption{(a) Ratio of the common halo mass, $M_{\rm h}$, to the total cluster mass,
$M_{\rm cl}$, as a function of time.
The circles represent model NB and
the solid line represents model FPB1.
(b) Evolution of the central density of the common halo component
(solid) and that of the galaxy component (dashed) for model FPB1.
The circles represent the central density of the common halo
for model NB.}
\end{figure}

\begin{figure}
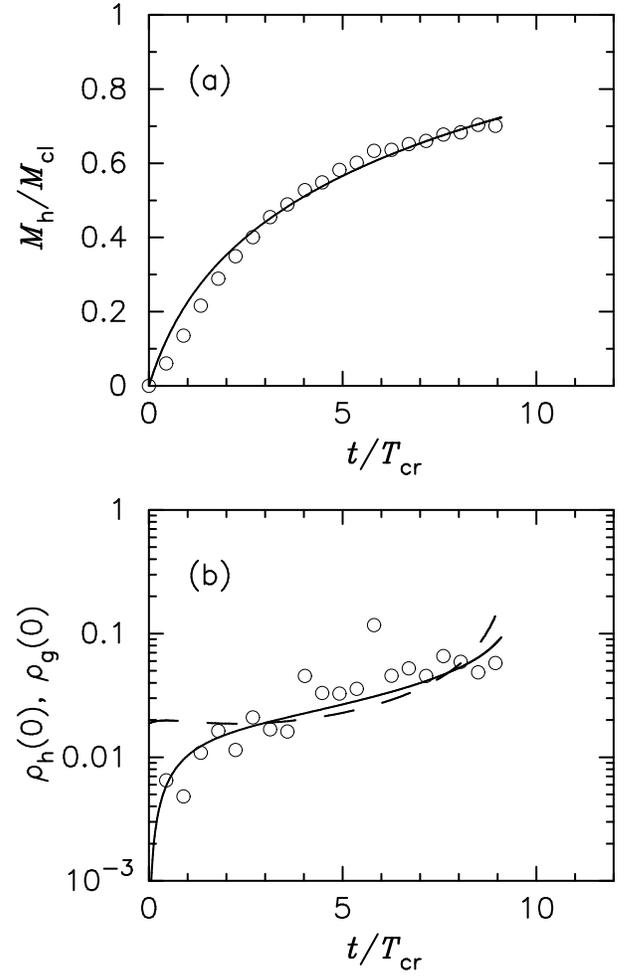

\begin{center}
\FigureFile(80mm,){fig3.ps}
\end{center}
\caption{Same as figure 2, but model FPC1 is compared with model NB.} 
\end{figure}

\clearpage

\begin{figure}
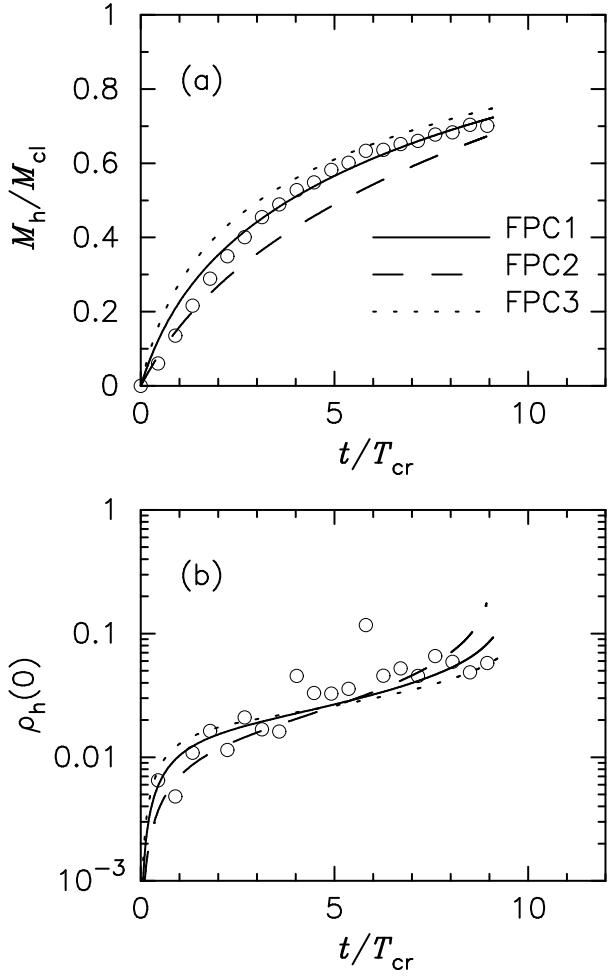

\begin{center}
\FigureFile(80mm,){fig4.ps}
\end{center}
\caption{Same as figure 2,
but the results for models FPC1, FPC2, and FPC3 are shown
by the solid, dashed, and dotted lines, respectively.
In (b) only the central density of the common halo is shown.}
\end{figure}

\begin{figure}
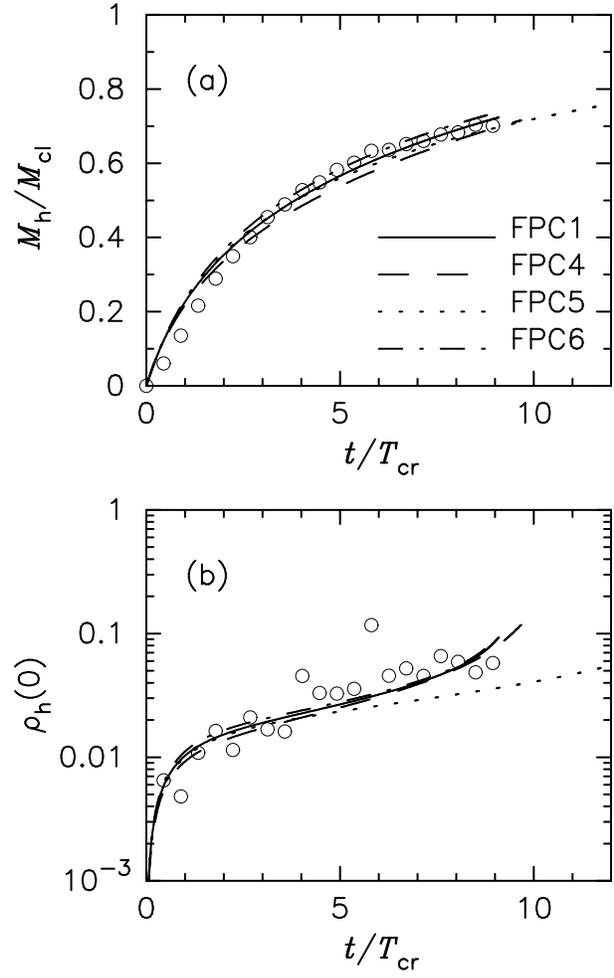

\begin{center}
\FigureFile(80mm,){fig5.ps}
\end{center}
\caption{Same as figure 4,
but the results for models FPC1, FPC4, FPC5, and FPC6 are shown by
the solid, dashed, dotted, and dash-dotted lines, respectively.}
\end{figure}

\clearpage

\onecolumn

\begin{figure}
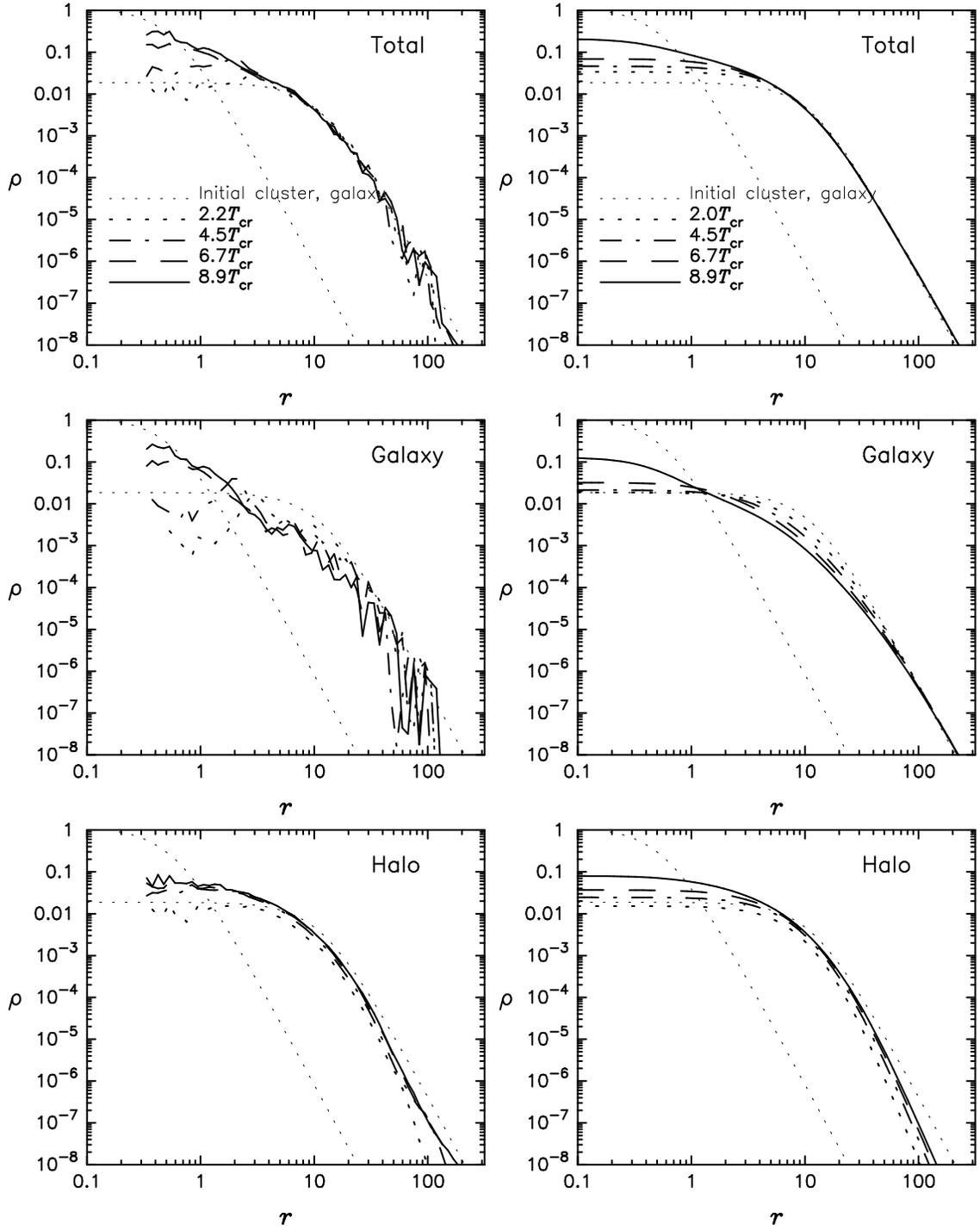

\begin{center}
\FigureFile(150mm,){fig6.ps}
\end{center}
\caption{Density profiles of models NB (left) and FPC1 (right) at selected
epochs. The top panels show the total density, and the middle and bottom panels
show the densities of the galaxy component and the common halo component,
respectively.
The density profile of the initial cluster
as well as the profile of an initial galaxy 
(here its center is placed at the cluster center) is plotted
with thin dotted lines.}
\end{figure}

\clearpage

\twocolumn

\begin{figure}
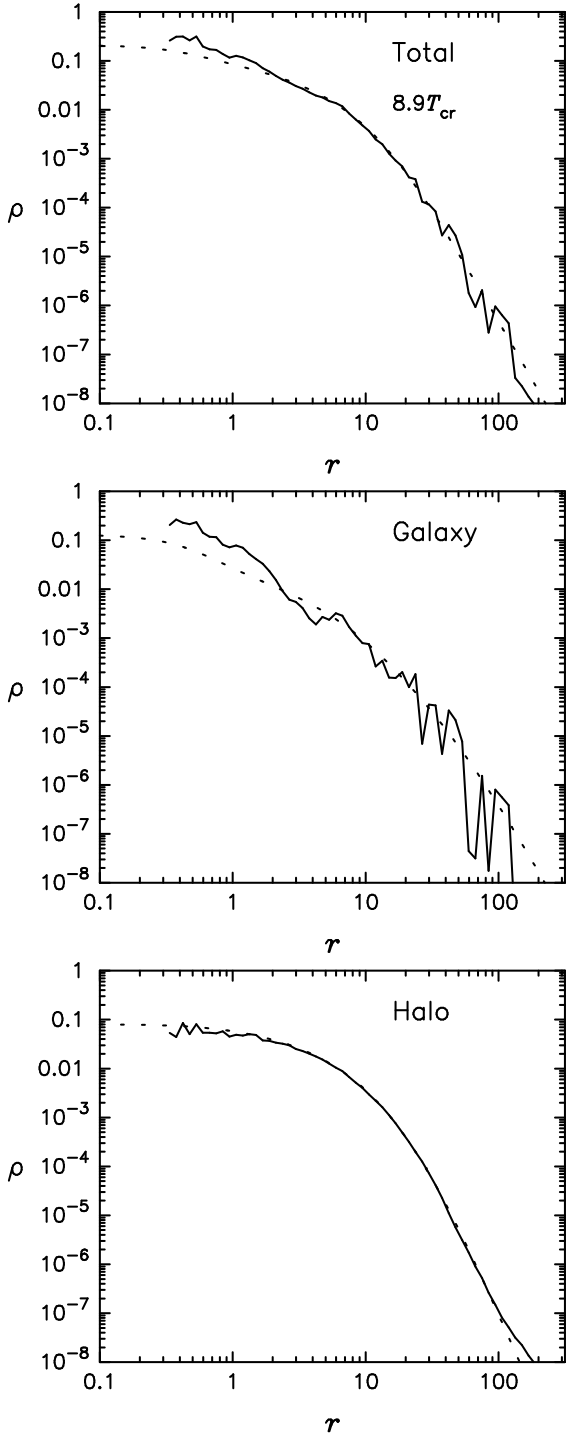

\begin{center}
\FigureFile(75mm,){fig7.ps}
\end{center}
\caption{Comparison of the density profiles of models NB (solid) and FPC1
(dotted) at $t=8.9\,T_{\rm cr}$. The top panel shows the total density,
and the middle and bottom panels
show the densities of the galaxy component and the common halo
component, respectively.
Note that the two models should be compared with caution for $r \ltsim 1$,
because in model NB the galaxies have sizes of $\sim 1$
and their internal structure is seen on the scale of $r \ltsim 1$.}
\end{figure}

\begin{figure}
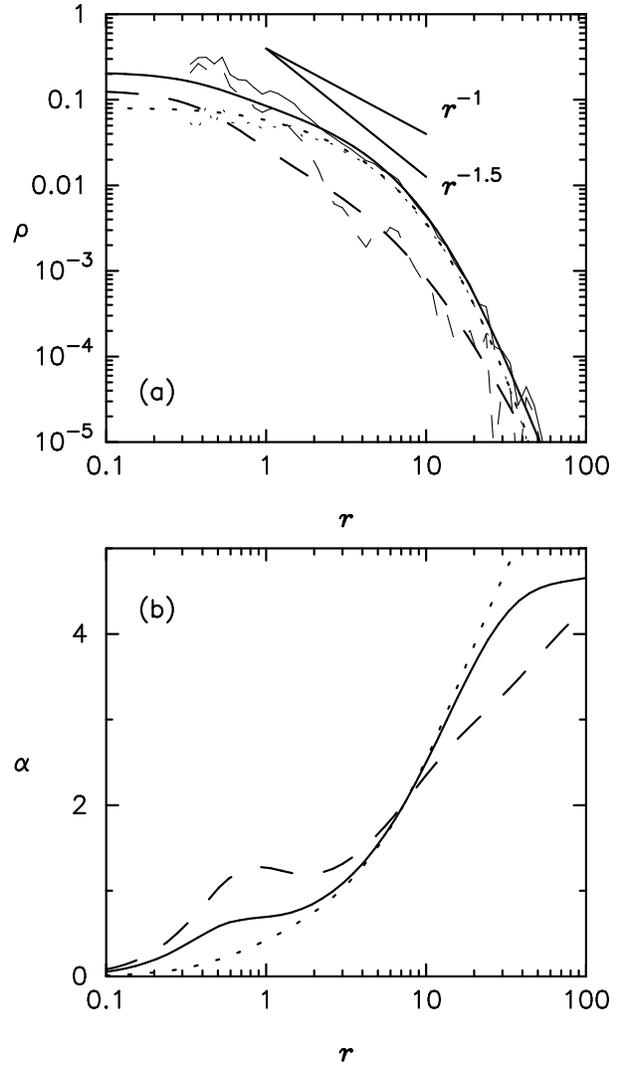

\begin{center}
\FigureFile(80mm,){fig8.ps}
\end{center}
\caption{(a) Density profile $\rho(r)$ for model FPC1 at $t=8.9\,T_{\rm cr}$.
The solid, dashed, and dotted lines represent
the total density, the galaxy component, and the common halo
component, respectively.
The density profiles for model NB
at $t=8.9\,T_{\rm cr}$ are also shown by thin lines.
(b) Logarithmic density gradient,
$\alpha (r) \equiv -d\ln \rho / d\ln r$, corresponding to the density profiles
for model FPC1 shown in panel (a).}
\end{figure}

\clearpage

\begin{figure}
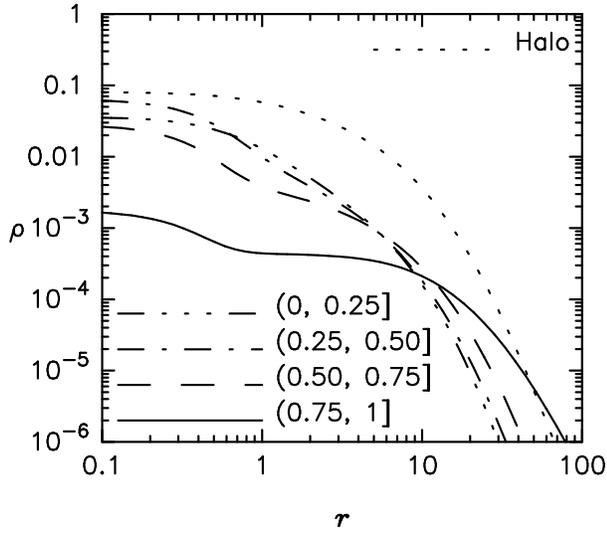

\begin{center}
\FigureFile(80mm,){fig9.ps}
\end{center}
\caption{Density profiles for the galaxies in four mass ranges,
$(0,0.25]$, $(0.25,0.50]$, $(0.50,0.75]$, and  $(0.75,1]$,
separately shown for model FPC1 at $t=8.9\,T_{\rm cr}$.
The dotted line is for the common halo.} 
\end{figure}

\begin{figure}
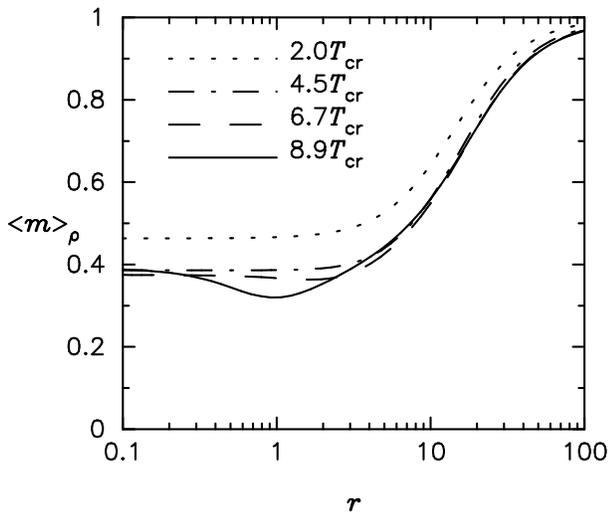

\begin{center}
\FigureFile(80mm,){fig10.ps}
\end{center}
\caption{Mean galaxy mass $\langle m \rangle_\rho$
as a function of radius for model FPC1 at selected epochs.}
\end{figure}

\begin{figure}
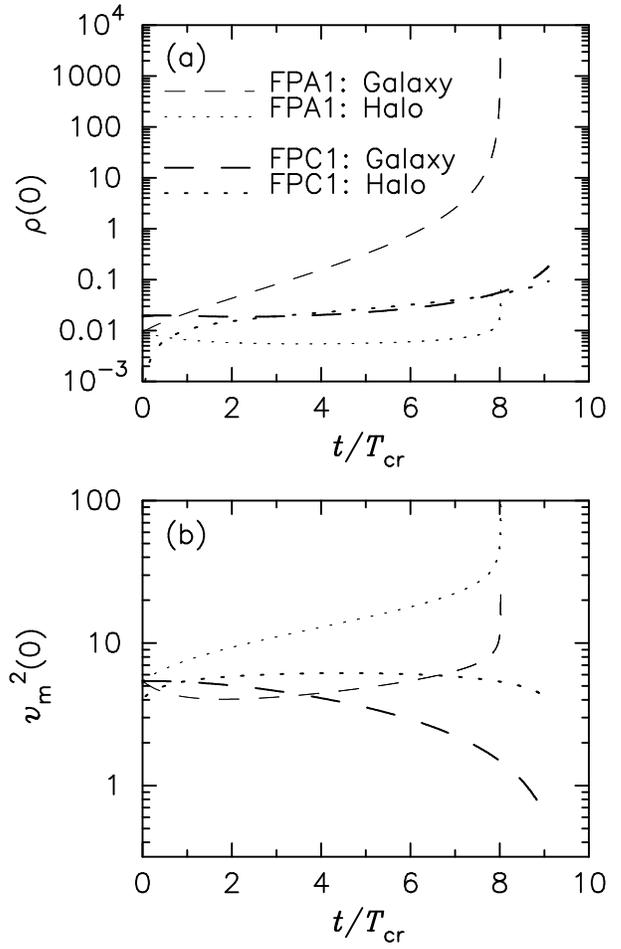

\begin{center}
\FigureFile(80mm,){fig11.ps}
\end{center}
\caption{(a) Evolution of the central densities of the galaxy component
(dashed) and the halo component (dotted) for models FPA1 and FPC1.
(b) Evolution of the central velocity dispersions of
the galaxy component (dashed) and  the halo component (dotted) for
models FPA1 and FPC1.}
\end{figure}

\clearpage

\begin{figure}
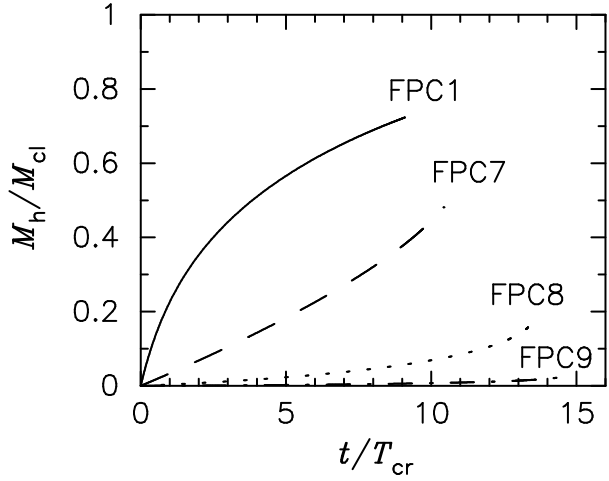

\begin{center}
\FigureFile(80mm,){fig12.ps}
\end{center}
\caption{Growth of the common halo mass for models FPC1, FPC7, FPC8, and FPC9.}
\end{figure}

\begin{figure}
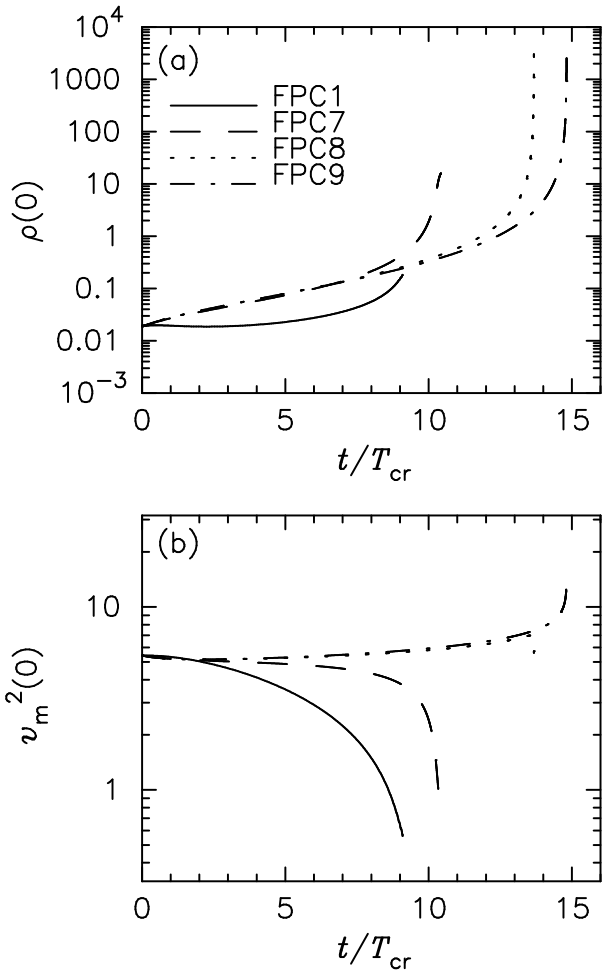

\begin{center}
\FigureFile(80mm,){fig13.ps}
\end{center}
\caption{(a) Evolution of the central density of the galaxy component for models
FPC1, FPC7, FPC8, and FPC9.
(b) Evolution of the central velocity dispersion of the galaxy
component for models FPC1, FPC7, FPC8, and FPC9.} 
\end{figure}

\begin{figure}
\begin{center}
\FigureFile(80mm,){fig14.ps}
\end{center}
\caption{(a) Evolution of the density profile for model FPC7.
The profiles at $t/T_{\rm cr} = 7.1$, 9.8, and 10.3 are plotted.
The solid, dashed, and dotted lines represent
the total density, the galaxy component, and the common halo
component, respectively.
(b) The profiles of the logarithmic density gradient, $\alpha$,
corresponding to the density profiles shown in panel (a).}
\end{figure}

\clearpage

\begin{figure}
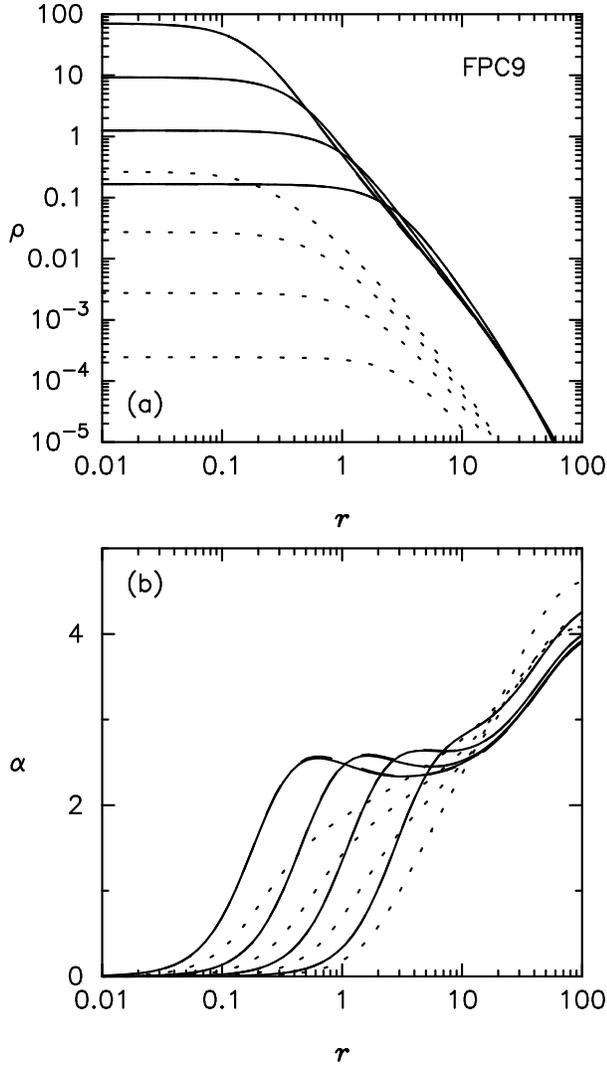

\begin{center}
\FigureFile(80mm,){fig15.ps}
\end{center}
\caption{Same as figure 14, but for model FPC9.
The profiles at $t/T_{\rm cr}=7.8$, 12.7, 14.3, and 14.7 are plotted.}
\end{figure}

\begin{figure}
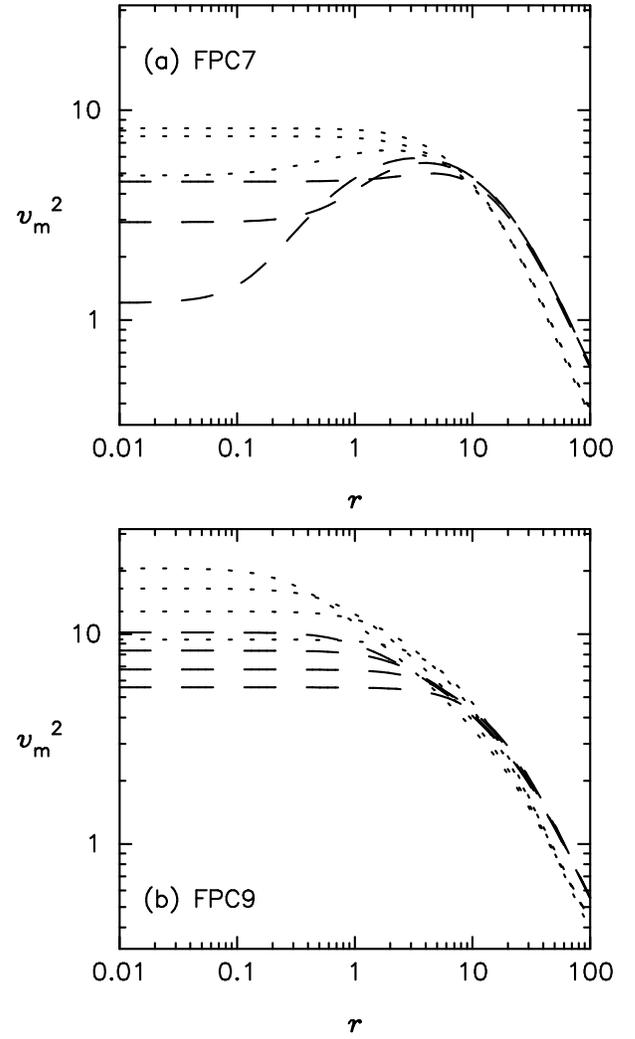

\begin{center}
\FigureFile(80mm,){fig16.ps}
\end{center}
\caption{(a) Evolution of the velocity dispersion profiles for model FPC7.
The profiles at the same epochs as in figure 14 are plotted.
The dashed and dotted lines represent the galaxy and common halo
components, respectively.
For the galaxy the mass-weighted mean velocity dispersion is plotted.
The central velocity dispersions of both components decrease with time.
(b) Same as (a), but for model FPC9.
The profiles at the same epochs as in figure 15 are plotted.
The central velocity dispersions of both components increase with time.}
\end{figure}

\end{document}